\documentclass[%
aps,pre,
superscriptaddress,
 amsmath,amssymb,
 reprint,%
]{revtex4-1}
\pdfoutput=1
\usepackage[T1]{fontenc}
\usepackage[utf8]{inputenc}
\usepackage{babel}
\usepackage{amsmath}
\usepackage{amssymb}
\usepackage{amsthm}
\usepackage{textcomp}
\usepackage{lmodern}
\usepackage{physics}
\usepackage{xcolor}
\usepackage{graphicx}
\usepackage[hidelinks,
colorlinks,
linkcolor=blue,
citecolor=blue,
urlcolor=blue,
]{hyperref}

\graphicspath{{figures/}}

\newcommand{\ee}{\mathrm{e}}
\newcommand{\ii}{\mathrm{i}}
\newcommand{\pp}{\mathrm{p}}
\newcommand{\nc}[1][]{n_{\mathrm{c}#1}}

\newcommand{\micro}{\text{\textmu}}

\usepackage[nosfdefault]{comicneue}
\newcommand{\Smilei}{{\comicneue Smilei}}

\begin{document}

\title[SRS with XUV attosecond pulses and pulse trains]%
{Stimulated-Raman-scattering amplification of attosecond XUV pulses
  with \\ pulse-train pumps and application to local in-depth
  plasma-density measurement} %
\thanks{\it Published under the terms of the
  \href{https://creativecommons.org/licenses/by/4.0/}{Creative Commons
    Attribution 4.0 International} license. Further distribution of
  this work must maintain attribution to the author(s) and the
  published article's title, journal citation, and DOI. }

\author{Andr\'eas~Sundstr\"om}%
\email{andsunds@chalmers.se}%
\affiliation{Department of Physics, Chalmers University of
  Technology, SE-412\,96 Gothenburg, Sweden.}%

\author{Mickael~Grech}%
\affiliation{LULI, CNRS, Sorbonne Universit\'{e}, CEA,
  \'{E}cole Polytechnique, Institut Polytechnique de Paris,
  F-91128 Palaiseau, France}%
\author{Istv\'{a}n~Pusztai}%
\affiliation{Department of Physics, Chalmers University of
  Technology, SE-412\,96 Gothenburg, Sweden.}%
\author{Caterina~Riconda}%
\affiliation{LULI, Sorbonne Universit\'{e}, CNRS, CEA,
  \'{E}cole Polytechnique, Institut Polytechnique de Paris,
  F-75252 Paris, France}%

\date{\today}

\begin{abstract}
We present a scheme for amplifying an extreme-ultraviolet (XUV) seed
isolated attosecond pulse via stimulated Raman scattering of a
pulse-train pump. At sufficient seed and pump intensity, the
amplification is nonlinear, and the amplitude of the seed pulse can
reach that of the pump, one order of magnitude higher than the initial
seed amplitude. In the linear amplification regime, we find that the
spectral signature of the pump pulse train is imprinted on the
spectrum of the amplified seed pulse. Since the spectral signature is
imprinted with its frequency downshifted by the plasma frequency, it
is possible to deduce the electron density in the region of
interaction. This region can be of micrometer length scale
longitudinally. By varying the delay between the seed and the pump,
this scheme provides a local electron-density measurement inside
solid-density plasmas that cannot be probed with optical frequencies,
with micrometer resolution.
\\[1em]
DOI:\href{https://doi.org/10.1103/PhysRevE.106.045208}{\tt
  10.1103/PhysRevE.106.045208}
\end{abstract}

\maketitle


\section{Introduction}

With high-order harmonic generation in gas targets (gas-HHG)~%
\cite{Ferray-etal_JPB1988,Paul-etal_Sci2001}, %
it is possible to generate isolated attosecond pulses (IAPs) in the
extreme-ultraviolet (XUV) regime~%
\cite{Christov-etal_PRL1997,Hentschel-etal_Nat2001,Itatani-etal_PRL2002,Xue-etal_SciAdv2020}. %
Such pulses can be used as probes to study processes on the attosecond
timescale~\cite{Calegari-etal_JPB2016}, e.g., electron wave-function
dynamics in atoms and molecules~%
\cite{Kling-Vrakking_AnnuRevPhysChem2008,Calegari-etal_Sci2014,Nisoli-etal_ChemRev2017},
as well as plasma diagnostics~%
\cite{Hergott-etal_LasPartBeams2001,Doboz-etal_PRL2005,Williams-etal_PoP2013,Koliyadu-etal_Photonics2017,AttoDispersion2022}. %
However, these pulses are relatively low energy and will, in general,
not be able to significantly drive a plasma in an XUV-pump--XUV-probe
experiment~\cite{Tzallas-etal_NatPhys2011,Takahashi-etal_NatCommun2013}.

Another XUV source is high-order harmonic generation in laser--solid
interactions~\cite{Gordienko-etal_PRL2004} (solid-HHG), via either
surface plasma waves~%
\cite{Brugge-etal_PRL2012,Fedeli-etal_ApplPhysLett2017}, %
coherent synchrotron emission~%
\cite{Brugge-Pukhov_PoP2010,Mikhailova-etal_PRL2012,Dromey-etal_NatPhys2012}, %
the relativistic electronic spring~%
\cite{Gonoskov-etal_PRE2011,Blackburn-etal_PRA2018}, %
or the relativistically oscillating mirror mechanism~%
\cite{Gibbon_PRL1996,Teubner-Gibbon_RevModPhys2009,Thaury-Quere_JPhysB2010,Vincenti_PRL2019,Fedeli-etal_PRL2021,Yi_PRL2021}. %
The latter has the potential to provide XUV pulses with very
high intensities, potentially up to
$I\sim10^{25}{\rm\,W/cm^{2}}$~\cite{Vincenti_PRL2019}.
With these methods, a train of high-intensity attosecond XUV pulses
can be generated with sufficiently high field strength to
significantly affect the electrons and drive collective processes in a
plasma.
There are also methods for generating IAPs with solid-HHG~%
\cite{Wheeler-etal_NatPhot2012,Vincenti-Quere_PRL2012,Hammond-Etal_Natphot2016,Mikhailova-etal_PRL2012,Heissler-etal_PRL2012,Kando-etal_QuantBeamSci2018,Jahn-etal_Optica2019}. %

In this paper we propose the use of stimulated Raman back-scattering
(SRS)~%
\cite{Maier-etal_PRL1966,Ping-etal_PRE2000,Cheng-etal_PRL2005,Ren-etal_NatPhys2007,Trines-etal_PRL2011,Trines-etal_SciRep2020} %
to amplify an IAP seed pulse with a pulse-train pump. While the
three-wave-coupling system of equations for SRS is fairly well
understood for pseudo-monochromatic pulses, its applicability to the
broad-spectrum IAP seeds and pulse-train pumps considered here is not
obvious. We therefore explore the possibility of this approach via
particle-in-cell (PIC) simulations, and present parametric studies of
the SRS amplification with respect to initial pump and seed amplitudes
as well as pulse-train length, plasma temperature, and the effect of
Coulomb collisions. The use of a pulse-train pump can be of interest
for smaller laser laboratories to produce high-intensity IAPs, as the other
pump-pulse alternative would be x-ray free-electron lasers (XFELs),
which are only found at a limited number of large-scale facilities.

Since SRS relies on the presence of electron-plasma waves induced by
the pump and seed laser pulses, the efficacy of the amplification is
highly dependent on both the pump and seed normalized
amplitudes~\cite{Trines-etal_SciRep2020}, $a_0$ and $a_1$,
respectively. Our simulations indicate that efficient nonlinear
amplification, without severe temporal stretching of the seed pulse,
requires $a_1\gtrsim0.02$, which is almost within reach for gas-HHG,
with reported amplitudes of
$a_1\simeq0.01$~\cite{Xue-etal_SciAdv2020}.

An interesting application of this XUV--SRS scheme is for in-depth,
local electron-density diagnostics.
Because high-density plasmas are opaque to optical frequencies,
light-based probing methods have to be in the XUV or x-ray range of
wavelengths. Previous experiments have utilized XUV
transmission~\cite{Hergott-etal_LasPartBeams2001,Doboz-etal_PRL2005}
to determine electron densities on ${\sim}100{\rm\,fs}$
timescales. Furthermore, XUV wave-front
sensing~\cite{Williams-etal_PoP2013} and dispersion of XUV
pulses~\cite{AttoDispersion2022} have been proposed to diagnose
solid-density laser-generated plasmas on the timescale of the XUV
pulse.  However, these techniques all produce line-integrated
measurements along the path of the probe pulse. By instead utilizing
the locality of the SRS interaction, the method proposed in this paper
has the potential to probe the electron density, inside the plasma, with
micrometer-scale longitudinal resolution by varying the delay between
the pump and seed.

The proposed local electron-density diagnostics method is based on the
spectral imprinting of the spectral fringes of the pump pulse train
onto the broad-spectrum seed pulse. This technique has previously been
proposed~\cite{Jang-etal_ApplPhysLett2008,Cho-etal_ApplPhysLett2014,Song-etal_PPCF2016},
and demonstrated experimentally~\cite{Vieux-etal_ApplPhysLett2013} to probe
laser-generated plasmas in gas-jet targets using SRS of infrared
lasers.
Besides the possibility of probing solid-density plasmas at
significantly higher spatial and temporal resolution employing XUV
frequencies, our proposed use of a pulse-train pump has an additional
advantage through an increased spectral resolution. Since a pulse
train has a spectrum consisting of several prominent peaks (depending
on the separation of the pulses in the train) and the SRS coupling is
strongest at a frequency shift equal to the plasma frequency
$\Delta\omega=\omega_{\pp}$, these spectral peaks will be imprinted
onto the spectrum of the seed pulse. The amplified seed pulse will
therefore have a very distinct spectral signature, which can be
compared with the original pulse-train pump to determine the local
plasma frequency $\omega_{\pp}$ and thereby the electron density.
Owing to the narrow spectral peaks, we estimate the uncertainty of
the density measurement using our proposed setup to be as small as
$2.5\,\%$.
%
Note that this spectral imprinting is most efficient in the linear SRS
amplification regime, i.e., for $a_1\lesssim0.01$, which is currently
realizable.

\section{Interaction regimes}
\label{sec:scales}

Consider a pump and a counterpropagating seed in a plasma with
frequencies $\omega_0$ and $\omega_1$, respectively, fulfilling the
phase matching condition $\omega_0-\omega_1=\omega_{\pp}$, where
$\omega_{\pp}$ is the plasma frequency; the pulses have normalized
amplitudes $a_i= eE_i/(m_{\ee}c\omega_i)$ (for $i=0,1$) with
$E_i=\sqrt{2I_i/c\epsilon_0}$ and where $e$ is the elementary charge,
$m_\ee$ is the electron mass, $c$ is the speed of light, $\epsilon_0$ is
the permittivity of free space, and $I_i$ is the intensity of the
respective beams. Following \citet{Edwards-etal_PRE2017}, the
linear-regime SRS growth rate $\varGamma$ for the seed pulse inside a
plasma with electron density $n_{\ee}$ is given by
\begin{equation}
\frac{\varGamma}{\omega_0}=\frac{1}{2}\qty[4\tilde{\varGamma}^{2}
+(\tilde{\nu}_{3}-\tilde{\nu}_{1})^{2}]^{1/2}
-\frac{\tilde{\nu}_{3}+\tilde{\nu}_{1}}{2},
\end{equation}
where 
\begin{equation}\label{eq:Gamma-tilde}
\tilde{\varGamma} = 
\frac{a_0c\,(|k_0|+|k_1|)\,(n_{\ee}/\nc[,0])^{1/4}}
{4\sqrt{\omega_0\omega_1}}
\end{equation}
is the undamped SRS growth rate with pump and seed wavenumbers $k_0$
and $k_1$, respectively, and $\nc[,0]=\epsilon_0m_{\ee}\omega_0^2/e^2$
is the critical density associated with $\omega_0$;
$\tilde{\nu}_{3}=\tilde{\nu}_{\rm Lnd}+\tilde{\nu}_{\ee\ii}/4$ and
$\tilde{\nu}_{1}=(n_{\ee}/\nc[,0])\,\tilde{\nu}_{\ee\ii}\omega_{0}/\omega_{1}$
using the normalized collisional and Landau damping rates
\begin{equation}
\tilde{\nu}_{\ee\ii}=\frac{2\sqrt{2}}{3\sqrt{\pi n_{\ee}/\nc[,0]}}
\frac{e^2 \omega_0 \ln{\varLambda}}{4\pi\epsilon_{0}m_{\ee}c^3 q_{\ee}^{3/2}}
\end{equation}
and
\begin{equation}
\tilde{\nu}_{\rm Lnd}=\frac{\sqrt{\pi n_{\ee}/\nc[,0]}}{(2q_{\ee})^{3/2}}
\,\exp\!\big({-}1/q_{\ee}\big),
\end{equation}
respectively, where $q_{\ee}=4T_{\ee}/(m_{\ee}c^2 n_{\ee}/\nc[,0])$,
and the Coulomb logarithm is taken as
$\ln\varLambda=\ln(12\pi\lambda_{\rm D}^2n_{\ee}/Z_{\ii})$ with the
electron temperature $T_{\ee}$, Debye length
$\lambda_{\rm D}=\sqrt{\epsilon_{0}T_{\ee}/(n_{\ee}e^2)}$, %
and ion charge number $Z_{\ii}$.

In this paper we study pump amplitudes ranging from $a_0=0.05$ to
$a_0=0.2$, which results in growth rates on the order of
$\varGamma/\omega_0\sim0.023{-}0.1$ at $n_{\ee}\approx0.6\,\nc[,0]$.
As discussed by \citet{Edwards-etal_PRE2017} for wavelengths above
$10 {\rm\,nm}$ and for the pump amplitudes considered here,
amplification by SRS should be more favorable than stimulated
Brillouin scattering (SBS), or in some cases the difference will be
marginal. However, this theory assumes quasimonochromatic pulses,
which is not the case in our study. We have also verified, by
complementary simulations (not shown here), that indeed the SBS
amplification is not efficient for our setup and parameter range. As
it is expected that the observed growth rates should be lower due to
the ``gaps'' between pulses in the pump pulse train, the faster time
scales associated with SRS are essential. Furthermore, due to the short
pulse durations, the energy in the pump and seed pulses is spread
over a broad spectral range, which is expected to differently affect
the growth rates of the spectral components of the seed. %
To our knowledge, there is no developed fully spectral theory of the
SRS with broad-band pulses, but we expect that the growth rate of each
spectral component is approximately proportional to the spectral
density $|\hat{E}|$ of the pump~--~similar to the $a_0$ dependence in
\eqref{eq:Gamma-tilde}~--~where $\hat{E}$ is the Fourier transform of
the electric field.

It has been pointed out that in order to have efficient SRS
amplification, it is useful to start with a seed that has initial
amplitude and duration close to the optimal condition of nonlinear
amplification~\cite{Trines-etal_SciRep2020}, namely,
$a_1 \tau_1 \omega_0 (n_{\ee}/\nc[,0] )^{1/4} \approx 6.8$, where
$a_1$ and $\tau_1$ are the amplitude and duration, respectively, of the
seed pulse once the nonlinear regime has set in, with $a_1$ being
comparable to $a_0$. Typical seed pulses in the XUV range will be only
a few cycles and will interact in relatively high density plasmas,
of the order of $0.1\nc[,0]{-}0.2\nc[,0]$.  In order to fulfill the
condition above, we would need a very large initial seed amplitude
$a_1=0.5$, which is beyond the parameter range of interest for this
paper. Nonetheless, for sufficiently large pump intensity
($a_0\gtrsim0.2$) we do enter into a nonlinear regime of amplification
as will be discussed below.

\section{Simulation setup}
\label{sec:sim}

We used the \Smilei{} PIC code~\cite{Smilei-paper} to simulate the
interaction between a seed pulse and a pump pulse train in a plasma in
one dimension (1D). The simulations were performed in various box
sizes, all with a spatial resolution of $\Delta{x}\simeq0.7{\rm\,nm}$,
at least 50 times smaller than the laser wavelengths used in this
study.
The seed pulse considered here had a central wavelength of
$\lambda_1=50{\rm\,nm}$ and the Gaussian temporal envelope with a
(field-amplitude) full-width-at-half-maximum (FWHM) duration of four
cycles. The plasma was chosen to be a pure hydrogen plasma at
densities $0.1\nc[,1]$ (Sec.~\ref{sec:non-lin}) and $0.2\nc[,1]$
(Sec.~\ref{sec:dens-diag}), where $\nc[,1]$ is the critical density
associated with the seed central frequency $\omega_{1}$. The plasma
was modeled using $1000$ and $800$ particles per cell for the
electrons and protons, respectively. The initial electron and ion
temperatures were $T_{\ee}=100{\rm\,eV}$ and $T_{\ii}=1{\rm\,eV}$,
respectively.
The pulse trains considered had central wavelengths of
$\lambda_0=38{\rm\,nm}$ (Sec.~\ref{sec:non-lin}) and
$\lambda_0=36{\rm\,nm}$ (Sec.~\ref{sec:dens-diag}); %
the individual pulses in the train were also of four-cycle FWHM
duration~--~based on their own central wavelength. The distance
between pulses in the train was $400{\rm\,nm}$.

We have also performed additional simulations to investigate the
effect of electron temperature and Coulomb collisions on the SRS as
well as on the spontaneous Raman scattering (RS) which occurs for the
higher-intensity pumps. This was done through a scan in electron
temperature from $50{\rm\,eV}$ to $500{\rm\,eV}$, both with and
without binary Coulomb collisions~\cite{Perez-etal_PoP2012} enabled.
Because the spontaneous RS is seeded by random density fluctuations,
it is sensitive to numerical noise, and we have used a higher
resolution in order to accurately capture the spontaneous RS %
\footnote{ The spatial resolution was doubled,
  $\Delta{x}\simeq0.34{\rm\,nm}$, and the number of particles per cell
  was increased to 8000 per cell and per species, i.e., in total 16
  times the number of particles compared with the nominal
  simulations. We deemed that the simulation results of the
  spontaneous RS were sufficiently converged, when the energy
  difference between 4000 and 8000 particles per cell was smaller than
  10\,\%.}. %
We note, however, that the strength of the spontaneous RS in our study
is relatively small compared with the SRS, and that the difference in
energy contained in the spontaneous RS between the nominal and
high-resolution simulation is ${\lesssim}25\%$, which amounts to
${\lesssim}2.5\%$ of the energy in the SRS-amplified pulse in all
cases. The nominal-resolution simulations therefore give a good
representation of the behavior of the SRS mechanism, even if they
might overestimate spontaneous RS.

We note that the use of 1D PIC simulations is well justified under
these conditions. The fastest-growing transverse instability one can
expect, the filamentation instability~%
\cite{Max-etal_PRL1974,Weber-etal_PRL2013}, grows over timescales of
hundreds of femtoseconds and submicrometer transverse scales, much
larger than those considered in this paper.

\section{Results and discussion}
\label{sec:res}

We first discuss the amplification of an IAP (Sec.~\ref{sec:non-lin}),
considering in particular the effect of the lengths of pump pulse
trains. Most efficient amplification, without severely extending the
duration of the amplified seed pulse, is shown to occur in the
non-linear regime.
In contrast, in the linear SRS regime (Sec.~\ref{sec:dens-diag}), the
clear spectral signature of the pump pulse train is imprinted onto the
seed pulse with a frequency shift equal to the local plasma frequency.
By comparing the spectra of the amplified and pump pulses, this
technique can be used to deduce the electron density inside the
plasma, with micrometer resolution.

\subsection{XUV-pulse amplification}
\label{sec:non-lin}

\begin{figure}
\centering
\includegraphics{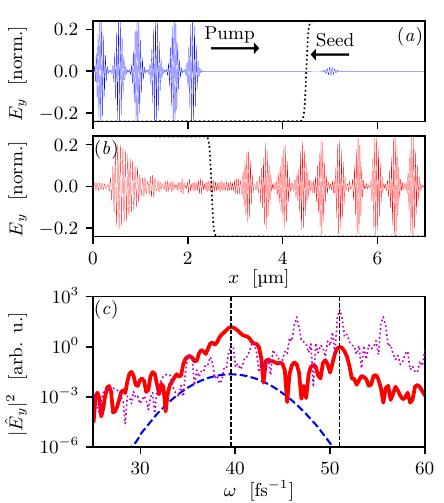}
\caption{Amplification of an IAP by an XUV pulse train via nonlinear
  SRS, with initial pump and seed amplitudes $a_0=0.2$ and $a_1=0.02$,
  respectively. (a) and (b) Real-space field [relativistically
  normalized (norm.) with the central frequency of the seed]
  before~(a) and after~(b) the seed--pump interaction. The dotted
  black curves indicate the spatial filtering envelopes used to
  distinguish the spectra of the seed or amplified pulse. (c)~Spectra
  of the amplified (solid red curve) and initial seed pulse (dashed
  blue curve), as well as the pump pulse train after interaction with
  the seed (dotted magenta curve).  The vertical dashed lines indicate
  the spectral maxima of the pump and amplified seed pulse. }
\label{fig:SRS-ampl}
\end{figure}

Figure~\ref{fig:SRS-ampl} shows the amplification of a seed pulse with
initial amplitude $a_1=0.02$; Figs.~\ref{fig:SRS-ampl}(a) and
\ref{fig:SRS-ampl}(b) show the pulse before and after, respectively,
interaction with the pump pulse train consisting of $N=10$ pulses with
wavelength $\lambda_0=38{\rm\,nm}$. The dotted black curves in
Figs.~\ref{fig:SRS-ampl}(a) and \ref{fig:SRS-ampl}(b) indicate the
spatial filtering used to obtain the spectra of the seed pulse shown
in Fig.~\ref{fig:SRS-ampl}(c), where the dashed blue curve represents
the spectrum of the initial seed pulse and the solid red curve
represents that of the amplified seed pulse; the dotted magenta curve
shows the spectrum of the pump after SRS has occurred. The vertical
dashed lines show the spectral maxima of the pump and amplified seed
pulses; the corresponding frequency downshift $\Delta\omega$ agrees
very well with the plasma frequency $\omega_\pp$ associated with the
electron density $n_{\ee}=0.1\nc[,1]$ used here.

The quality of the amplification can broadly be captured by two
parameters: the energy gain %
$\mathcal{G}= \mathcal{E}_1^{\rm final}/\mathcal{E}_1^{\rm init}$ %
and amplified-pulse duration $\tau_{95\%}$ measured as the time span
that contains 95\,\% of the energy in the pulse.
By integrating the spatially filtered spectra in
Fig.~\ref{fig:SRS-ampl}(c), we can calculate the energy
$\mathcal{E}_1$ of the seed pulse before and after amplification,
which gives an energy gain of $\mathcal{G}\simeq236$ %
in this case. %
Next, we find the duration of the amplified pulse to be
$\tau_{95\%}=3.0{\rm\,fs}$; in comparison, the initial seed-pulse
duration is $\tau_{95\%}^{\rm(init)}\!\!\approx0.8{\rm\,fs}$.

In order to gauge the efficacy of the pulse-train pump compared with
flat-top (constant-intensity) pulses, we run two simulations with two
different flat-top pump pulses. The first such pulse was chosen to
contain the same total energy as the $N=10$ pulse train, resulting in
a duration of $3.5{\rm\,fs}$. The second flat-top pump had a duration
of $10.6{\rm\,fs}$, which is the same duration as the pulse
train~--~but containing approximately 2.9 times more energy. The
energy gain $\mathcal{G}$ and amplified-pulse duration $\tau_{95\%}$
obtained from the pulse-train pump and the two flat-top pumps are
summarized in Table~\ref{tab:comparison}.
The lower energy gain $\mathcal{G}$ observed with the same-energy
flat-top pump might indicate that the seed pulse requires some fixed
time in the linear-growth regime, which makes up a larger proportion of
the same-energy pump than in the pulse train. At the same time, the
short duration $\tau_{95\%}$ and observed amplified-pulse spectrum
(not shown here) suggest that the SRS interaction eventually reaches
the nonlinear regime, with resulting pulse compression.
The same-duration flat-top pump results in a significantly larger
energy gain, as expected, while still having a significantly shorter
duration than the pulse-train pump; both of these observations are
expected since the higher energy in this pump allows for both more
energy transfer and a more nonlinear interaction.
It also appears that the flat-top pumps result in somewhat less
spontaneous RS and other noise, which is likely due to the narrower
spectrum of the flat-top pulse envelopes.

\begin{table}
\caption{Comparison of the energy gains $\mathcal{G}$ and
  amplified-pulse duration $\tau_{95\%}$ between the nominal
  pulse-train pump with $N=10$ train length (as shown in
  Fig.~\ref{fig:SRS-ampl}) and two cases with flat-top pumps: either
  with the same total energy or with the same duration as the nominal
  pulse train. }
\label{tab:comparison}
\begin{tabular}{lcc}\hline\hline
  Pump type& \hspace{-1em}Energy gain $\mathcal{G}$ \ &  Duration $\tau_{95\%}$ (fs) \\\hline
  Pulse train & 236 & 3.0 \\
  Flat top (same energy) & 83 & 0.9 \\
  Flat top (same duration) & 330 & 1.5 \\\hline\hline
\end{tabular}
\end{table}

\begin{figure}
\centering
\includegraphics{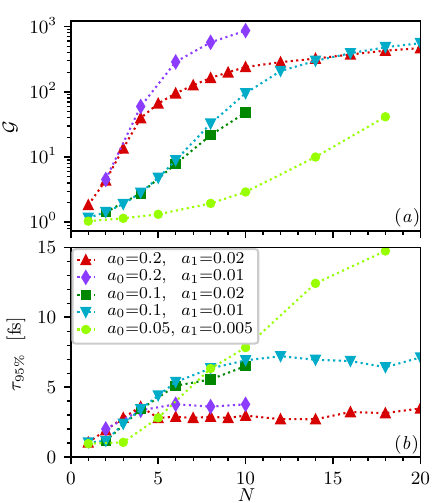}
\caption{Relative energy gain $\mathcal{G}$ (a) and amplified-pulse
  duration $\tau_{95\%}$ (b) as functions of the number of pulses $N$ in
  the pump train. The seed and pump pulses are both of four-cycle FWHM
  duration with seed and pump wavelengths $\lambda_1=50{\rm\,nm}$ and
  $\lambda_0=38{\rm\,nm}$, respectively. The plasma density is
  $n_{\ee}=0.1\,\nc[,1]$. }
\label{fig:train-length}
\end{figure}

We have also performed similar calculations for various pulse-train
lengths $N$ with different combinations of pump and seed amplitudes,
$a_0$ and $a_1$, respectively, which are shown in
Fig.~\ref{fig:train-length}(a). In the cases displayed, the gain rises
rapidly with the number of pump pulses for small $N$. However, at a
certain point the gain saturates at approximately
$\mathcal{G}\sim(a_0/a_1)^2$, which roughly corresponds to the seed
pulse reaching the amplitude of the pump. Extending the pump pulse
train beyond this saturation point mostly results in a broadening of
the pulse envelope and an increased background of spontaneous Raman
scattering~(RS).
The unwanted spontaneous RS is especially prominent in the cases with
$a_0=0.2$ and $a_1=0.02$~--~due to the stronger nonlinear effects at
higher pump amplitudes. Already at $N=10$, there is noticeable
spontaneous RS, as seen by the noise floor on both sides of the
amplified seed pulse in Fig.~\ref{fig:SRS-ampl}(b).

The stronger nonlinearity present at higher pump and seed amplitudes
is, however, also key to producing short amplified pulses.  In
Fig.~\ref{fig:train-length}(b), we see that the shortest
amplified-pulse durations $\tau_{95\%}\simeq3{\rm\,fs}$ are produced
with $a_0=0.2$ (with $N$ between 5 and 10). %
For the $[a_0,a_1]=[0.2,0.02]$ case (red triangles), we also observe a
decrease in $\tau_{95\%}$ starting at $N=5$, which indicates the onset
of pulse compression observed in strongly nonlinear SRS
\cite{Trines-etal_NatPhys2011}.  For the same case, we also see a more
irregular behavior after $N=10$, which is likely due to the higher
spontaneous RS in the longer simulation box used for the $N>10$ cases,
as discussed in the previous paragraph.

\begin{figure}
\centering
\includegraphics{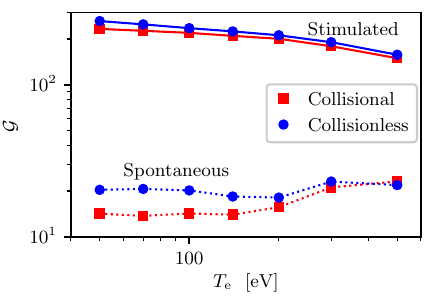}
\caption{Relative energy gain $\mathcal{G}$ of the seed pulse (solid
  curves)~--~stimulated scattering~--~and energy of the spontaneous RS
  (dotted curves), as functions of electron temperature $T_{\ee}$, with
  (red squares) and without collisions (blue circles). All cases shown
  here are with $a_0=0.2$ and $a_1=0.02$ and with a $N=10$ pulse-train
  pump.  Both the stimulated and spontaneous energy gains are measured
  relative to the energy of the initial seed pulse. }
\label{fig:Temp-scan}
\end{figure}

We continue by investigating the saturation of the relative energy gain by
varying the initial electron temperature, which affects the strength
of Landau damping, as well as using a seed pulse with $a_1=a_0=0.2$,
to study whether electron-wave breaking may be causing the saturation.
These simulations were performed with significantly higher resolution
in order to accurately resolve the spontaneous RS, as discussed in
Sec.~\ref{sec:sim}; corresponding results are shown in
fig.~\ref{fig:Temp-scan}. 
When the electron temperature is varied between $T_{\ee}=50{\rm\,eV}$
and $T_{\ee}=500{\rm\,eV}$, there is no clear trend in the strength of the
spontaneous RS in the collisionless cases, while the desired
stimulated RS is reduced at higher temperatures. Interestingly,
including collisions reduces the spontaneous RS somewhat while at the
same time only impacting the stimulated energy gain by
${\lesssim}10{\,\%}$. However, in foil targets the higher ion charge
at solid density may result in more significant collisional
effects~\cite{ElectronPaper2020,IonPaper2020}. The effect of
collisions decreases with temperature, both in stimulated RS and in
spontaneous RS. At $T_{\ee}=500{\rm\,eV}$, the spontaneous RS is even
slightly higher with collisions than without; this is within the
noise margin of the PIC simulations.

\begin{figure}
\centering
\includegraphics{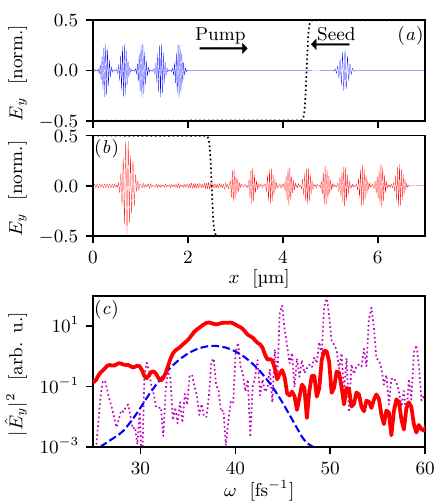}
\caption{Amplification of an IAP by an XUV pulse train with the same
  amplitude seed and pump $a_0=a_1=0.2$. (a) and (b) Real-space field
  (relativistically normalized with the central frequency of the seed)
  before~(a) and after~(b) the seed--pump interaction. The dotted
  black curvess indicate the spatial filtering envelopes used to
  distinguish the spectra of the seed or amplified pulse. (c)~Spectra
  of the amplified (solid red curve) and initial seed pulse (dashed
  blue curve), as well as the pump pulse train after interaction with
  the seed (dotted magenta curve).  }
\label{fig:same-ampl}
\end{figure}

Next, in a series of simulations with $a_1=a_0=0.2$, we varied the
number of pulses in the pump train from $N=2$ to $N=10$ to study the
effect of electron-wave breaking. %
The result of the $N=10$ case is shown in Fig.~\ref{fig:same-ampl},
where we in Fig.~\ref{fig:same-ampl}(b) see that the amplitude of the amplified pulse is
almost $2.5$ times higher than the initial seed amplitude, without any
significant increase in pulse duration~--~the amplified-pulse duration
is $\tau_{95\%}\approx1.0{\rm\,fs}$, compared with the initial seed
duration of $\tau_{95\%}^{\rm(init)}\!\approx0.8{\rm\,fs}$. %
While the electron phase space (not shown) in the $a_1=a_0=0.2$ cases
shows clear tendencies of electron-wave breaking~--~which occurs when
the amplitude of the electron wave is sufficiently large that the
fastest electrons significantly outrun the wave itself~--~we still
find that the absolute energy gain %
$\Delta\mathcal{E}_1 %
=\mathcal{E}_1^{\rm final}-\mathcal{E}_1^{\rm init}$ %
is approximately twice that of the $[a_0,a_1]=[0.2,\,0.02]$ case.
Thus, since there is no deterioration of the energy transfer for
$a_1=a_0=0.2$, we may conclude that electron-wave breaking is not
significantly affecting the nonlinear amplification in the cases
considered here.

\subsection{Linear SRS as a local density diagnostic}
\label{sec:dens-diag}

\begin{figure}
\centering
\includegraphics{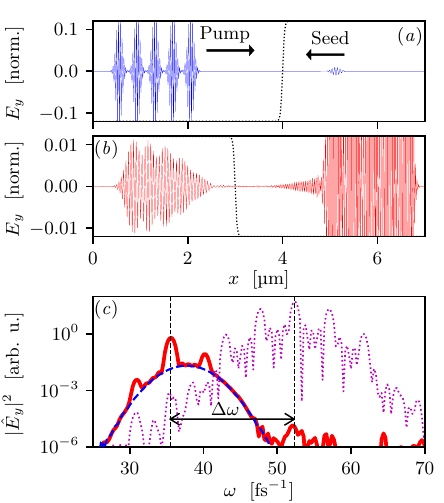}
\caption{Illustration of the spectral imprinting from linear SRS, with
  initial pump and seed amplitudes $a_0=0.2$ and $a_1=0.02$,
  respectively. (a) and (b) Real-space field (relativistically
  normalized with the central frequency of the seed) before~(a) and
  after~(b) the seed--pump interaction. The dotted black curves
  indicate the spatial filtering envelopes used to distinguish the
  spectra of the seed pulse. (c)~Spectra of the amplified (solid red
  curve) and initial seed pulse (dashed blue line), as well as the
  pump after interaction with the seed (dotted magenta curve). The
  vertical dashed lines indicate the spectral maxima of the pump and
  amplified seed pulse, separated by a frequency shift of
  $\Delta\omega\approx16.8{\rm\,fs^{-1}}$.  }
\label{fig:dens-diag}
\end{figure}

Although the relative energy gain is lower and the amplified-pulse
duration is longer in the linear compared with the nonlinear regime,
the same mechanism that causes these undesirable amplification
properties can instead be exploited for measuring the electron density
inside a plasma. Figure~\ref{fig:dens-diag} shows the results of the
interaction between an $a_0=0.1$ pump pulse train and an $a_1=0.01$
seed pulse with central wavelengths (in vacuum) of
$\lambda_0=36{\rm\,nm}$ and $\lambda_1=50{\rm\,nm}$,
respectively. Figures~\ref{fig:dens-diag}(a) and \ref{fig:dens-diag}(b) show the pulses before and after
interaction, respectively. We find that the amplitude of the seed was
not greatly affected by the SRS; however, the pulse duration
increased to $\tau_{95\%}\approx4.6{\rm\,fs}$~--~practically to that
of the full pump pulse train.

The cause for this result is elucidated by the spectra shown
in Fig.~\ref{fig:dens-diag}(c). The amplified pulse has acquired three
distinct peaks (solid red curve) on top of the initial spectrum (dashed
blue curve). These peaks correspond to a downshift by $\omega_{\pp}$
of the three highest peaks of the pump spectrum (dotted magenta curve). 
Spectrally, the amplified seed pulse has become more similar to a
pulse train, which is also seen in the real-space shape of the
envelope [Fig.~\ref{fig:dens-diag}(b)]. Owing to the frequency shift,
the spectral peaks of the amplified seed pulse are still
distinguishable from reflections of the pump pulse train.

The linear nature of the low-amplitude SRS means that the
amplification of each spectral component of the seed is approximately
proportional to the spectral density $|\hat{E}_y^{(0)}(\omega)|$ of
the pump (downshifted by $\omega_{\pp}$). We therefore get an
imprinting of the rather distinct spectral signature of the pump pulse
train onto the amplified pulse, which allows us to determine the size
of the frequency shift $\Delta\omega=16.83{\rm\,fs^{-1}}$, as
indicated by the two vertical dashed lines in
Fig.~\ref{fig:dens-diag}(c). We find that the downshift agrees
very well with the electron density $n_{\ee}=0.2\nc[,1]$ used in this
simulation, which has a corresponding plasma frequency of
$\omega_{\pp}=16.85{\rm\,fs^{-1}}$. In practice, however, the accuracy
of the measured frequency shift is limited by the width of the
spectral peaks, which in this case is approximately
$\pm\,0.8{\rm\,fs^{-1}}$, giving an uncertainty in the density
measurement of less than ${\pm}2.5\,\%$.
If the length, $N$, of the pulse train is increased, the
spectral width of each peak will be reduced, thus increasing the
accuracy~--~albeit at the cost of lower spatiotemporal resolution.

Also note that the broad-spectrum nature of the seed and the pump is
crucial for this density diagnostic method to be effective.  Because
the spectral amplification also depends on the spectral density of the
seed, $|\hat{E}_y^{(1)}(\omega)|$, the seed-pulse spectrum should be
sufficiently broad to capture the imprinting of an as-wide-as-possible
section of the pump spectrum, even when the central frequencies are
not perfectly matched~--~as is the case in the simulation shown in
Fig.~\ref{fig:dens-diag}.
If the seed and pump had narrow spectra, then the frequency difference
between the seed and the pump would have to be swept, in order to find the
largest amplification, before the value of $\omega_{\pp}$ could be
established, which could be challenging experimentally. %
Of course, the method is still limited in what plasma densities can be
measured~--~based on which seed and pump central frequencies are
available~--~even with a broadband seed and pump, but the range of
densities that can be probed increases with the seed and pump
bandwidths.

Next, because the spectral imprinting only occurs while the seed and
the pump interact, the probed frequency shift corresponds to the plasma
frequency only in the interaction region, the size of which is
approximately half the length of the pump pulse train.
While the linear regime operates with relatively low amplitude pulses,
it should be noted that if the pump amplitude is below
$a_0\lesssim0.05$, the SRS growth rate becomes low~--~even if the seed
amplitude is increased to $a_1=a_0$~--~and a longer pump pulse train
is needed, thus increasing the interaction region.
By varying the pump--seed delay, it is possible to choose where in
the plasma to probe the density.  This method therefore has the
potential to probe the electron density at micrometer longitudinal
resolution, at depth inside the plasma. Depending on the transverse
size of the focus, the probed volume can be as small as a few tens of
$\rm\micro{m}^3$.

\section{Conclusions}
\label{sec:concl}

We have presented a scheme for amplifying isolated attosecond XUV
pulses based on stimulated Raman scattering (SRS). %
In order to efficiently amplify the attosecond seed pulse without
severely extending its duration, the seed and pump must be
sufficiently strong to enter the nonlinear regime. In particular, we
find that the relativistic amplitude of the seed should be
$a_1\gtrsim0.02$ with a pump of amplitude $a_0=0.2$. %

In the linear SRS regime, the broad spectrum of the attosecond seed is
imprinted with the peaked spectrum of the pump pulse train.
Because the imprinting of the pump spectrum occurs with a frequency
downshift equal to the local plasma frequency
$\omega_{\pp}\propto{}n_{\ee}^{1/2}$, the local plasma density can be
deduced by comparing the spectra of the pump and amplified seed
pulses.
The advantage of this in-depth density-measurement method is
that, depending on the length of the pump pulse train, the
longitudinal spatial resolution can be in the micrometer scale,
allowing for local density diagnostics of solid-density plasmas inside
the plasma.

\begin{acknowledgments}
The authors are grateful for fruitful discussions with P.\
Eng-Johnsson, A.\ Gonoskov, and T.\ F\"{u}l\"{o}p. %
This project has received funding from the Knut and Alice Wallenberg
Foundation (Grand no.\ KAW\,2020.0111). %
A.S.\ gratefully acknowledges the support from Adlerbertska
forskningsstiftelsen.
The computations were enabled by resources provided by the Swedish
National Infrastructure for Computing (SNIC), partially funded by the
Swedish Research Council through grant agreement no.\ 2018-05973.
\end{acknowledgments}

\bibliography{references}

\begin{thebibliography}{56}%
\makeatletter
\providecommand \@ifxundefined [1]{%
 \@ifx{#1\undefined}
}%
\providecommand \@ifnum [1]{%
 \ifnum #1\expandafter \@firstoftwo
 \else \expandafter \@secondoftwo
 \fi
}%
\providecommand \@ifx [1]{%
 \ifx #1\expandafter \@firstoftwo
 \else \expandafter \@secondoftwo
 \fi
}%
\providecommand \natexlab [1]{#1}%
\providecommand \enquote  [1]{``#1''}%
\providecommand \bibnamefont  [1]{#1}%
\providecommand \bibfnamefont [1]{#1}%
\providecommand \citenamefont [1]{#1}%
\providecommand \href@noop [0]{\@secondoftwo}%
\providecommand \href [0]{\begingroup \@sanitize@url \@href}%
\providecommand \@href[1]{\@@startlink{#1}\@@href}%
\providecommand \@@href[1]{\endgroup#1\@@endlink}%
\providecommand \@sanitize@url [0]{\catcode `\\12\catcode `\$12\catcode
  `\&12\catcode `\#12\catcode `\^12\catcode `\_12\catcode `\%12\relax}%
\providecommand \@@startlink[1]{}%
\providecommand \@@endlink[0]{}%
\providecommand \url  [0]{\begingroup\@sanitize@url \@url }%
\providecommand \@url [1]{\endgroup\@href {#1}{\urlprefix }}%
\providecommand \urlprefix  [0]{URL }%
\providecommand \Eprint [0]{\href }%
\providecommand \doibase [0]{http://dx.doi.org/}%
\providecommand \selectlanguage [0]{\@gobble}%
\providecommand \bibinfo  [0]{\@secondoftwo}%
\providecommand \bibfield  [0]{\@secondoftwo}%
\providecommand \translation [1]{[#1]}%
\providecommand \BibitemOpen [0]{}%
\providecommand \bibitemStop [0]{}%
\providecommand \bibitemNoStop [0]{.\EOS\space}%
\providecommand \EOS [0]{\spacefactor3000\relax}%
\providecommand \BibitemShut  [1]{\csname bibitem#1\endcsname}%
\let\auto@bib@innerbib\@empty
\bibitem [{\citenamefont {Ferray}\ \emph {et~al.}(1988)\citenamefont {Ferray},
  \citenamefont {L'Huillier}, \citenamefont {Li}, \citenamefont {Lompre},
  \citenamefont {Mainfray},\ and\ \citenamefont {Manus}}]{Ferray-etal_JPB1988}%
  \BibitemOpen
  \bibfield  {author} {\bibinfo {author} {\bibfnamefont {M.}~\bibnamefont
  {Ferray}}, \bibinfo {author} {\bibfnamefont {A.}~\bibnamefont {L'Huillier}},
  \bibinfo {author} {\bibfnamefont {X.~F.}\ \bibnamefont {Li}}, \bibinfo
  {author} {\bibfnamefont {L.~A.}\ \bibnamefont {Lompre}}, \bibinfo {author}
  {\bibfnamefont {G.}~\bibnamefont {Mainfray}}, \ and\ \bibinfo {author}
  {\bibfnamefont {C.}~\bibnamefont {Manus}},\ }\href {\doibase
  10.1088/0953-4075/21/3/001} {\bibfield  {journal} {\bibinfo  {journal}
  {Journal of Physics B: Atomic, Molecular and Optical Physics}\ }\textbf
  {\bibinfo {volume} {21}},\ \bibinfo {pages} {L31} (\bibinfo {year}
  {1988})}\BibitemShut {NoStop}%
\bibitem [{\citenamefont {Paul}\ \emph {et~al.}(2001)\citenamefont {Paul},
  \citenamefont {Toma}, \citenamefont {Breger}, \citenamefont {Mullot},
  \citenamefont {Aug\'e}, \citenamefont {{Balcou, {Ph}.}}, \citenamefont
  {Muller},\ and\ \citenamefont {Agostini}}]{Paul-etal_Sci2001}%
  \BibitemOpen
  \bibfield  {author} {\bibinfo {author} {\bibfnamefont {P.~M.}\ \bibnamefont
  {Paul}}, \bibinfo {author} {\bibfnamefont {E.~S.}\ \bibnamefont {Toma}},
  \bibinfo {author} {\bibfnamefont {P.}~\bibnamefont {Breger}}, \bibinfo
  {author} {\bibfnamefont {G.}~\bibnamefont {Mullot}}, \bibinfo {author}
  {\bibfnamefont {F.}~\bibnamefont {Aug\'e}}, \bibinfo {author} {\bibnamefont
  {{Balcou, {Ph}.}}}, \bibinfo {author} {\bibfnamefont {H.~G.}\ \bibnamefont
  {Muller}}, \ and\ \bibinfo {author} {\bibfnamefont {P.}~\bibnamefont
  {Agostini}},\ }\href {\doibase 10.1126/science.1059413} {\bibfield  {journal}
  {\bibinfo  {journal} {Science}\ }\textbf {\bibinfo {volume} {292}},\ \bibinfo
  {pages} {1689} (\bibinfo {year} {2001})}\BibitemShut {NoStop}%
\bibitem [{\citenamefont {Christov}\ \emph {et~al.}(1997)\citenamefont
  {Christov}, \citenamefont {Murnane},\ and\ \citenamefont
  {Kapteyn}}]{Christov-etal_PRL1997}%
  \BibitemOpen
  \bibfield  {author} {\bibinfo {author} {\bibfnamefont {I.~P.}\ \bibnamefont
  {Christov}}, \bibinfo {author} {\bibfnamefont {M.~M.}\ \bibnamefont
  {Murnane}}, \ and\ \bibinfo {author} {\bibfnamefont {H.~C.}\ \bibnamefont
  {Kapteyn}},\ }\href {\doibase 10.1103/PhysRevLett.78.1251} {\bibfield
  {journal} {\bibinfo  {journal} {Phys. Rev. Lett.}\ }\textbf {\bibinfo
  {volume} {78}},\ \bibinfo {pages} {1251} (\bibinfo {year}
  {1997})}\BibitemShut {NoStop}%
\bibitem [{\citenamefont {Hentschel}\ \emph {et~al.}(2001)\citenamefont
  {Hentschel}, \citenamefont {Kienberger}, \citenamefont {Spielmann},
  \citenamefont {Reider}, \citenamefont {Milosevic}, \citenamefont {Brabec},
  \citenamefont {Corkum}, \citenamefont {Heinzmann}, \citenamefont {Drescher},\
  and\ \citenamefont {Krausz}}]{Hentschel-etal_Nat2001}%
  \BibitemOpen
  \bibfield  {author} {\bibinfo {author} {\bibfnamefont {M.}~\bibnamefont
  {Hentschel}}, \bibinfo {author} {\bibfnamefont {R.}~\bibnamefont
  {Kienberger}}, \bibinfo {author} {\bibfnamefont {C.}~\bibnamefont
  {Spielmann}}, \bibinfo {author} {\bibfnamefont {G.~A.}\ \bibnamefont
  {Reider}}, \bibinfo {author} {\bibfnamefont {N.}~\bibnamefont {Milosevic}},
  \bibinfo {author} {\bibfnamefont {T.}~\bibnamefont {Brabec}}, \bibinfo
  {author} {\bibfnamefont {P.}~\bibnamefont {Corkum}}, \bibinfo {author}
  {\bibfnamefont {U.}~\bibnamefont {Heinzmann}}, \bibinfo {author}
  {\bibfnamefont {M.}~\bibnamefont {Drescher}}, \ and\ \bibinfo {author}
  {\bibfnamefont {F.}~\bibnamefont {Krausz}},\ }\href {\doibase
  10.1038/35107000} {\bibfield  {journal} {\bibinfo  {journal} {Nature}\
  }\textbf {\bibinfo {volume} {414}},\ \bibinfo {pages} {509} (\bibinfo {year}
  {2001})}\BibitemShut {NoStop}%
\bibitem [{\citenamefont {Itatani}\ \emph {et~al.}(2002)\citenamefont
  {Itatani}, \citenamefont {Qu\'er\'e}, \citenamefont {Yudin}, \citenamefont
  {Ivanov}, \citenamefont {Krausz},\ and\ \citenamefont
  {Corkum}}]{Itatani-etal_PRL2002}%
  \BibitemOpen
  \bibfield  {author} {\bibinfo {author} {\bibfnamefont {J.}~\bibnamefont
  {Itatani}}, \bibinfo {author} {\bibfnamefont {F.}~\bibnamefont {Qu\'er\'e}},
  \bibinfo {author} {\bibfnamefont {G.~L.}\ \bibnamefont {Yudin}}, \bibinfo
  {author} {\bibfnamefont {M.~Y.}\ \bibnamefont {Ivanov}}, \bibinfo {author}
  {\bibfnamefont {F.}~\bibnamefont {Krausz}}, \ and\ \bibinfo {author}
  {\bibfnamefont {P.~B.}\ \bibnamefont {Corkum}},\ }\href {\doibase
  10.1103/PhysRevLett.88.173903} {\bibfield  {journal} {\bibinfo  {journal}
  {Phys. Rev. Lett.}\ }\textbf {\bibinfo {volume} {88}},\ \bibinfo {pages}
  {173903} (\bibinfo {year} {2002})}\BibitemShut {NoStop}%
\bibitem [{\citenamefont {Xue}\ \emph {et~al.}(2020)\citenamefont {Xue},
  \citenamefont {Tamaru}, \citenamefont {Fu}, \citenamefont {Yuan},
  \citenamefont {Lan}, \citenamefont {M{\"u}cke}, \citenamefont {Suda},
  \citenamefont {Midorikawa},\ and\ \citenamefont
  {Takahashi}}]{Xue-etal_SciAdv2020}%
  \BibitemOpen
  \bibfield  {author} {\bibinfo {author} {\bibfnamefont {B.}~\bibnamefont
  {Xue}}, \bibinfo {author} {\bibfnamefont {Y.}~\bibnamefont {Tamaru}},
  \bibinfo {author} {\bibfnamefont {Y.}~\bibnamefont {Fu}}, \bibinfo {author}
  {\bibfnamefont {H.}~\bibnamefont {Yuan}}, \bibinfo {author} {\bibfnamefont
  {P.}~\bibnamefont {Lan}}, \bibinfo {author} {\bibfnamefont {O.~D.}\
  \bibnamefont {M{\"u}cke}}, \bibinfo {author} {\bibfnamefont {A.}~\bibnamefont
  {Suda}}, \bibinfo {author} {\bibfnamefont {K.}~\bibnamefont {Midorikawa}}, \
  and\ \bibinfo {author} {\bibfnamefont {E.~J.}\ \bibnamefont {Takahashi}},\
  }\href {\doibase 10.1126/sciadv.aay2802} {\bibfield  {journal} {\bibinfo
  {journal} {Science Advances}\ }\textbf {\bibinfo {volume} {6}},\ \bibinfo
  {pages} {eaay2802} (\bibinfo {year} {2020})}\BibitemShut {NoStop}%
\bibitem [{\citenamefont {Calegari}\ \emph {et~al.}(2016)\citenamefont
  {Calegari}, \citenamefont {Sansone}, \citenamefont {Stagira}, \citenamefont
  {Vozzi},\ and\ \citenamefont {Nisoli}}]{Calegari-etal_JPB2016}%
  \BibitemOpen
  \bibfield  {author} {\bibinfo {author} {\bibfnamefont {F.}~\bibnamefont
  {Calegari}}, \bibinfo {author} {\bibfnamefont {G.}~\bibnamefont {Sansone}},
  \bibinfo {author} {\bibfnamefont {S.}~\bibnamefont {Stagira}}, \bibinfo
  {author} {\bibfnamefont {C.}~\bibnamefont {Vozzi}}, \ and\ \bibinfo {author}
  {\bibfnamefont {M.}~\bibnamefont {Nisoli}},\ }\href {\doibase
  10.1088/0953-4075/49/6/062001} {\bibfield  {journal} {\bibinfo  {journal}
  {Journal of Physics {B:} Atomic, Molecular and Optical Physics}\ }\textbf
  {\bibinfo {volume} {49}},\ \bibinfo {pages} {062001} (\bibinfo {year}
  {2016})}\BibitemShut {NoStop}%
\bibitem [{\citenamefont {Kling}\ and\ \citenamefont
  {Vrakking}(2008)}]{Kling-Vrakking_AnnuRevPhysChem2008}%
  \BibitemOpen
  \bibfield  {author} {\bibinfo {author} {\bibfnamefont {M.~F.}\ \bibnamefont
  {Kling}}\ and\ \bibinfo {author} {\bibfnamefont {M.~J.~J.}\ \bibnamefont
  {Vrakking}},\ }\href {\doibase 10.1146/annurev.physchem.59.032607.093532}
  {\bibfield  {journal} {\bibinfo  {journal} {Annu. Rev. Phys. Chem.}\ }\textbf
  {\bibinfo {volume} {59}},\ \bibinfo {pages} {463} (\bibinfo {year}
  {2008})}\BibitemShut {NoStop}%
\bibitem [{\citenamefont {Calegari}\ \emph {et~al.}(2014)\citenamefont
  {Calegari}, \citenamefont {Ayuso}, \citenamefont {Trabattoni}, \citenamefont
  {Belshaw}, \citenamefont {De~Camillis}, \citenamefont {Anumula},
  \citenamefont {Frassetto}, \citenamefont {Poletto}, \citenamefont {Palacios},
  \citenamefont {Decleva} \emph {et~al.}}]{Calegari-etal_Sci2014}%
  \BibitemOpen
  \bibfield  {author} {\bibinfo {author} {\bibfnamefont {F.}~\bibnamefont
  {Calegari}}, \bibinfo {author} {\bibfnamefont {D.}~\bibnamefont {Ayuso}},
  \bibinfo {author} {\bibfnamefont {A.}~\bibnamefont {Trabattoni}}, \bibinfo
  {author} {\bibfnamefont {L.}~\bibnamefont {Belshaw}}, \bibinfo {author}
  {\bibfnamefont {S.}~\bibnamefont {De~Camillis}}, \bibinfo {author}
  {\bibfnamefont {S.}~\bibnamefont {Anumula}}, \bibinfo {author} {\bibfnamefont
  {F.}~\bibnamefont {Frassetto}}, \bibinfo {author} {\bibfnamefont
  {L.}~\bibnamefont {Poletto}}, \bibinfo {author} {\bibfnamefont
  {A.}~\bibnamefont {Palacios}}, \bibinfo {author} {\bibfnamefont
  {P.}~\bibnamefont {Decleva}},  \emph {et~al.},\ }\href {\doibase
  10.1126/science.1254061} {\bibfield  {journal} {\bibinfo  {journal}
  {Science}\ }\textbf {\bibinfo {volume} {346}},\ \bibinfo {pages} {336}
  (\bibinfo {year} {2014})}\BibitemShut {NoStop}%
\bibitem [{\citenamefont {Nisoli}\ \emph {et~al.}(2017)\citenamefont {Nisoli},
  \citenamefont {Decleva}, \citenamefont {Calegari}, \citenamefont {Palacios},\
  and\ \citenamefont {Mart{\'\i}n}}]{Nisoli-etal_ChemRev2017}%
  \BibitemOpen
  \bibfield  {author} {\bibinfo {author} {\bibfnamefont {M.}~\bibnamefont
  {Nisoli}}, \bibinfo {author} {\bibfnamefont {P.}~\bibnamefont {Decleva}},
  \bibinfo {author} {\bibfnamefont {F.}~\bibnamefont {Calegari}}, \bibinfo
  {author} {\bibfnamefont {A.}~\bibnamefont {Palacios}}, \ and\ \bibinfo
  {author} {\bibfnamefont {F.}~\bibnamefont {Mart{\'\i}n}},\ }\href {\doibase
  10.1021/acs.chemrev.6b00453} {\bibfield  {journal} {\bibinfo  {journal}
  {Chemical reviews}\ }\textbf {\bibinfo {volume} {117}},\ \bibinfo {pages}
  {10760} (\bibinfo {year} {2017})}\BibitemShut {NoStop}%
\bibitem [{\citenamefont {Hergott}\ \emph {et~al.}(2001)\citenamefont
  {Hergott}, \citenamefont {Sali\`{e}res}, \citenamefont {Merdji},
  \citenamefont {le~D\'{e}roff}, \citenamefont {Carr\'{e}}, \citenamefont
  {Auguste}, \citenamefont {Monot}, \citenamefont {d'Oliveira}, \citenamefont
  {Descamps}, \citenamefont {Norin}, \citenamefont {Lyng\aa}, \citenamefont
  {L'Huillier}, \citenamefont {Wahlstr\"{o}m}, \citenamefont {Bellini},\ and\
  \citenamefont {Huller}}]{Hergott-etal_LasPartBeams2001}%
  \BibitemOpen
  \bibfield  {author} {\bibinfo {author} {\bibfnamefont {J.-F.}\ \bibnamefont
  {Hergott}}, \bibinfo {author} {\bibfnamefont {P.}~\bibnamefont
  {Sali\`{e}res}}, \bibinfo {author} {\bibfnamefont {H.}~\bibnamefont
  {Merdji}}, \bibinfo {author} {\bibfnamefont {L.}~\bibnamefont
  {le~D\'{e}roff}}, \bibinfo {author} {\bibfnamefont {B.}~\bibnamefont
  {Carr\'{e}}}, \bibinfo {author} {\bibfnamefont {T.}~\bibnamefont {Auguste}},
  \bibinfo {author} {\bibfnamefont {P.}~\bibnamefont {Monot}}, \bibinfo
  {author} {\bibfnamefont {P.}~\bibnamefont {d'Oliveira}}, \bibinfo {author}
  {\bibfnamefont {D.}~\bibnamefont {Descamps}}, \bibinfo {author}
  {\bibfnamefont {J.}~\bibnamefont {Norin}}, \bibinfo {author} {\bibfnamefont
  {C.}~\bibnamefont {Lyng\aa}}, \bibinfo {author} {\bibfnamefont
  {A.}~\bibnamefont {L'Huillier}}, \bibinfo {author} {\bibfnamefont {C.-G.}\
  \bibnamefont {Wahlstr\"{o}m}}, \bibinfo {author} {\bibfnamefont
  {M.}~\bibnamefont {Bellini}}, \ and\ \bibinfo {author} {\bibfnamefont
  {S.}~\bibnamefont {Huller}},\ }\href {\doibase 10.1017/S0263034601191056}
  {\bibfield  {journal} {\bibinfo  {journal} {Laser and Particle Beams}\
  }\textbf {\bibinfo {volume} {19}},\ \bibinfo {pages} {35–40} (\bibinfo
  {year} {2001})}\BibitemShut {NoStop}%
\bibitem [{\citenamefont {Dobosz}\ \emph {et~al.}(2005)\citenamefont {Dobosz},
  \citenamefont {Doumy}, \citenamefont {Stabile}, \citenamefont {D'Oliveira},
  \citenamefont {Monot}, \citenamefont {R\'eau}, \citenamefont {H\"uller},\
  and\ \citenamefont {Martin}}]{Doboz-etal_PRL2005}%
  \BibitemOpen
  \bibfield  {author} {\bibinfo {author} {\bibfnamefont {S.}~\bibnamefont
  {Dobosz}}, \bibinfo {author} {\bibfnamefont {G.}~\bibnamefont {Doumy}},
  \bibinfo {author} {\bibfnamefont {H.}~\bibnamefont {Stabile}}, \bibinfo
  {author} {\bibfnamefont {P.}~\bibnamefont {D'Oliveira}}, \bibinfo {author}
  {\bibfnamefont {P.}~\bibnamefont {Monot}}, \bibinfo {author} {\bibfnamefont
  {F.}~\bibnamefont {R\'eau}}, \bibinfo {author} {\bibfnamefont
  {S.}~\bibnamefont {H\"uller}}, \ and\ \bibinfo {author} {\bibfnamefont
  {P.}~\bibnamefont {Martin}},\ }\href {\doibase 10.1103/PhysRevLett.95.025001}
  {\bibfield  {journal} {\bibinfo  {journal} {Phys. Rev. Lett.}\ }\textbf
  {\bibinfo {volume} {95}},\ \bibinfo {pages} {025001} (\bibinfo {year}
  {2005})}\BibitemShut {NoStop}%
\bibitem [{\citenamefont {Williams}\ \emph {et~al.}(2013)\citenamefont
  {Williams}, \citenamefont {Chung}, \citenamefont {Vinko}, \citenamefont
  {K\"{u}nzel}, \citenamefont {Sardinha}, \citenamefont {{Zeitoun, {Ph.}}},\
  and\ \citenamefont {Fajardo}}]{Williams-etal_PoP2013}%
  \BibitemOpen
  \bibfield  {author} {\bibinfo {author} {\bibfnamefont {G.~O.}\ \bibnamefont
  {Williams}}, \bibinfo {author} {\bibfnamefont {H.-K.}\ \bibnamefont {Chung}},
  \bibinfo {author} {\bibfnamefont {S.~M.}\ \bibnamefont {Vinko}}, \bibinfo
  {author} {\bibfnamefont {S.}~\bibnamefont {K\"{u}nzel}}, \bibinfo {author}
  {\bibfnamefont {A.~B.}\ \bibnamefont {Sardinha}}, \bibinfo {author}
  {\bibnamefont {{Zeitoun, {Ph.}}}}, \ and\ \bibinfo {author} {\bibfnamefont
  {M.}~\bibnamefont {Fajardo}},\ }\href {\doibase 10.1063/1.4794964} {\bibfield
   {journal} {\bibinfo  {journal} {Physics of Plasmas}\ }\textbf {\bibinfo
  {volume} {20}},\ \bibinfo {pages} {042701} (\bibinfo {year}
  {2013})}\BibitemShut {NoStop}%
\bibitem [{\citenamefont {Koliyadu}\ \emph {et~al.}(2017)\citenamefont
  {Koliyadu}, \citenamefont {Künzel}, \citenamefont {Wodzinski}, \citenamefont
  {Keitel}, \citenamefont {Duarte}, \citenamefont {Williams}, \citenamefont
  {João}, \citenamefont {Pires}, \citenamefont {Hariton}, \citenamefont
  {Galletti}, \citenamefont {Gomes}, \citenamefont {Figueira}, \citenamefont
  {Dias}, \citenamefont {Lopes}, \citenamefont {Zeitoun}, \citenamefont
  {Plönjes},\ and\ \citenamefont {Fajardo}}]{Koliyadu-etal_Photonics2017}%
  \BibitemOpen
  \bibfield  {author} {\bibinfo {author} {\bibfnamefont {J.~C.~P.}\
  \bibnamefont {Koliyadu}}, \bibinfo {author} {\bibfnamefont {S.}~\bibnamefont
  {Künzel}}, \bibinfo {author} {\bibfnamefont {T.}~\bibnamefont {Wodzinski}},
  \bibinfo {author} {\bibfnamefont {B.}~\bibnamefont {Keitel}}, \bibinfo
  {author} {\bibfnamefont {J.}~\bibnamefont {Duarte}}, \bibinfo {author}
  {\bibfnamefont {G.~O.}\ \bibnamefont {Williams}}, \bibinfo {author}
  {\bibfnamefont {C.~P.}\ \bibnamefont {João}}, \bibinfo {author}
  {\bibfnamefont {H.}~\bibnamefont {Pires}}, \bibinfo {author} {\bibfnamefont
  {V.}~\bibnamefont {Hariton}}, \bibinfo {author} {\bibfnamefont
  {M.}~\bibnamefont {Galletti}}, \bibinfo {author} {\bibfnamefont
  {N.}~\bibnamefont {Gomes}}, \bibinfo {author} {\bibfnamefont
  {G.}~\bibnamefont {Figueira}}, \bibinfo {author} {\bibfnamefont {J.~M.}\
  \bibnamefont {Dias}}, \bibinfo {author} {\bibfnamefont {N.}~\bibnamefont
  {Lopes}}, \bibinfo {author} {\bibfnamefont {P.}~\bibnamefont {Zeitoun}},
  \bibinfo {author} {\bibfnamefont {E.}~\bibnamefont {Plönjes}}, \ and\
  \bibinfo {author} {\bibfnamefont {M.}~\bibnamefont {Fajardo}},\ }\href
  {\doibase 10.3390/photonics4020025} {\bibfield  {journal} {\bibinfo
  {journal} {Photonics}\ }\textbf {\bibinfo {volume} {4}},\ \bibinfo {pages}
  {25} (\bibinfo {year} {2017})}\BibitemShut {NoStop}%
\bibitem [{\citenamefont {Sundstr\"{o}m}\ \emph {et~al.}(2022)\citenamefont
  {Sundstr\"{o}m}, \citenamefont {Pusztai}, \citenamefont {Eng-Johnsson},\ and\
  \citenamefont {F\"{u}l\"{o}p}}]{AttoDispersion2022}%
  \BibitemOpen
  \bibfield  {author} {\bibinfo {author} {\bibfnamefont {A.}~\bibnamefont
  {Sundstr\"{o}m}}, \bibinfo {author} {\bibfnamefont {I.}~\bibnamefont
  {Pusztai}}, \bibinfo {author} {\bibfnamefont {P.}~\bibnamefont
  {Eng-Johnsson}}, \ and\ \bibinfo {author} {\bibfnamefont {T.}~\bibnamefont
  {F\"{u}l\"{o}p}},\ }\href {\doibase 10.1017/S0022377822000307} {\bibfield
  {journal} {\bibinfo  {journal} {Journal of Plasma Physics}\ }\textbf
  {\bibinfo {volume} {88}},\ \bibinfo {pages} {905880211} (\bibinfo {year}
  {2022})}\BibitemShut {NoStop}%
\bibitem [{\citenamefont {Tzallas}\ \emph {et~al.}(2011)\citenamefont
  {Tzallas}, \citenamefont {Skantzakis}, \citenamefont {Nikolopoulos},
  \citenamefont {Tsakiris},\ and\ \citenamefont
  {Charalambidis}}]{Tzallas-etal_NatPhys2011}%
  \BibitemOpen
  \bibfield  {author} {\bibinfo {author} {\bibfnamefont {P.}~\bibnamefont
  {Tzallas}}, \bibinfo {author} {\bibfnamefont {E.}~\bibnamefont {Skantzakis}},
  \bibinfo {author} {\bibfnamefont {L.}~\bibnamefont {Nikolopoulos}}, \bibinfo
  {author} {\bibfnamefont {G.~D.}\ \bibnamefont {Tsakiris}}, \ and\ \bibinfo
  {author} {\bibfnamefont {D.}~\bibnamefont {Charalambidis}},\ }\href {\doibase
  10.1038/nphys2033} {\bibfield  {journal} {\bibinfo  {journal} {Nature
  Physics}\ }\textbf {\bibinfo {volume} {7}},\ \bibinfo {pages} {781} (\bibinfo
  {year} {2011})}\BibitemShut {NoStop}%
\bibitem [{\citenamefont {Takahashi}\ \emph {et~al.}(2013)\citenamefont
  {Takahashi}, \citenamefont {Lan}, \citenamefont {M{\"u}cke}, \citenamefont
  {Nabekawa},\ and\ \citenamefont {Midorikawa}}]{Takahashi-etal_NatCommun2013}%
  \BibitemOpen
  \bibfield  {author} {\bibinfo {author} {\bibfnamefont {E.~J.}\ \bibnamefont
  {Takahashi}}, \bibinfo {author} {\bibfnamefont {P.}~\bibnamefont {Lan}},
  \bibinfo {author} {\bibfnamefont {O.~D.}\ \bibnamefont {M{\"u}cke}}, \bibinfo
  {author} {\bibfnamefont {Y.}~\bibnamefont {Nabekawa}}, \ and\ \bibinfo
  {author} {\bibfnamefont {K.}~\bibnamefont {Midorikawa}},\ }\href {\doibase
  10.1038/ncomms3691} {\bibfield  {journal} {\bibinfo  {journal} {Nature
  communications}\ }\textbf {\bibinfo {volume} {4}},\ \bibinfo {pages} {1}
  (\bibinfo {year} {2013})}\BibitemShut {NoStop}%
\bibitem [{\citenamefont {Gordienko}\ \emph {et~al.}(2004)\citenamefont
  {Gordienko}, \citenamefont {Pukhov}, \citenamefont {Shorokhov},\ and\
  \citenamefont {Baeva}}]{Gordienko-etal_PRL2004}%
  \BibitemOpen
  \bibfield  {author} {\bibinfo {author} {\bibfnamefont {S.}~\bibnamefont
  {Gordienko}}, \bibinfo {author} {\bibfnamefont {A.}~\bibnamefont {Pukhov}},
  \bibinfo {author} {\bibfnamefont {O.}~\bibnamefont {Shorokhov}}, \ and\
  \bibinfo {author} {\bibfnamefont {T.}~\bibnamefont {Baeva}},\ }\href
  {\doibase 10.1103/PhysRevLett.93.115002} {\bibfield  {journal} {\bibinfo
  {journal} {Phys. Rev. Lett.}\ }\textbf {\bibinfo {volume} {93}},\ \bibinfo
  {pages} {115002} (\bibinfo {year} {2004})}\BibitemShut {NoStop}%
\bibitem [{\citenamefont {an~der Br\"ugge}\ \emph {et~al.}(2012)\citenamefont
  {an~der Br\"ugge}, \citenamefont {Kumar}, \citenamefont {Pukhov},\ and\
  \citenamefont {R\"odel}}]{Brugge-etal_PRL2012}%
  \BibitemOpen
  \bibfield  {author} {\bibinfo {author} {\bibfnamefont {D.}~\bibnamefont
  {an~der Br\"ugge}}, \bibinfo {author} {\bibfnamefont {N.}~\bibnamefont
  {Kumar}}, \bibinfo {author} {\bibfnamefont {A.}~\bibnamefont {Pukhov}}, \
  and\ \bibinfo {author} {\bibfnamefont {C.}~\bibnamefont {R\"odel}},\ }\href
  {\doibase 10.1103/PhysRevLett.108.125002} {\bibfield  {journal} {\bibinfo
  {journal} {Phys. Rev. Lett.}\ }\textbf {\bibinfo {volume} {108}},\ \bibinfo
  {pages} {125002} (\bibinfo {year} {2012})}\BibitemShut {NoStop}%
\bibitem [{\citenamefont {Fedeli}\ \emph {et~al.}(2017)\citenamefont {Fedeli},
  \citenamefont {Sgattoni}, \citenamefont {Cantono},\ and\ \citenamefont
  {Macchi}}]{Fedeli-etal_ApplPhysLett2017}%
  \BibitemOpen
  \bibfield  {author} {\bibinfo {author} {\bibfnamefont {L.}~\bibnamefont
  {Fedeli}}, \bibinfo {author} {\bibfnamefont {A.}~\bibnamefont {Sgattoni}},
  \bibinfo {author} {\bibfnamefont {G.}~\bibnamefont {Cantono}}, \ and\
  \bibinfo {author} {\bibfnamefont {A.}~\bibnamefont {Macchi}},\ }\href
  {\doibase 10.1063/1.4975365} {\bibfield  {journal} {\bibinfo  {journal}
  {Applied Physics Letters}\ }\textbf {\bibinfo {volume} {110}},\ \bibinfo
  {pages} {051103} (\bibinfo {year} {2017})}\BibitemShut {NoStop}%
\bibitem [{\citenamefont {an~der Br\"{u}gge}\ and\ \citenamefont
  {Pukhov}(2010)}]{Brugge-Pukhov_PoP2010}%
  \BibitemOpen
  \bibfield  {author} {\bibinfo {author} {\bibfnamefont {D.}~\bibnamefont
  {an~der Br\"{u}gge}}\ and\ \bibinfo {author} {\bibfnamefont {A.}~\bibnamefont
  {Pukhov}},\ }\href {\doibase 10.1063/1.3353050} {\bibfield  {journal}
  {\bibinfo  {journal} {Physics of Plasmas}\ }\textbf {\bibinfo {volume}
  {17}},\ \bibinfo {pages} {033110} (\bibinfo {year} {2010})}\BibitemShut
  {NoStop}%
\bibitem [{\citenamefont {Mikhailova}\ \emph {et~al.}(2012)\citenamefont
  {Mikhailova}, \citenamefont {Fedorov}, \citenamefont {Karpowicz},
  \citenamefont {Gibbon}, \citenamefont {Platonenko}, \citenamefont
  {Zheltikov},\ and\ \citenamefont {Krausz}}]{Mikhailova-etal_PRL2012}%
  \BibitemOpen
  \bibfield  {author} {\bibinfo {author} {\bibfnamefont {J.~M.}\ \bibnamefont
  {Mikhailova}}, \bibinfo {author} {\bibfnamefont {M.~V.}\ \bibnamefont
  {Fedorov}}, \bibinfo {author} {\bibfnamefont {N.}~\bibnamefont {Karpowicz}},
  \bibinfo {author} {\bibfnamefont {P.}~\bibnamefont {Gibbon}}, \bibinfo
  {author} {\bibfnamefont {V.~T.}\ \bibnamefont {Platonenko}}, \bibinfo
  {author} {\bibfnamefont {A.~M.}\ \bibnamefont {Zheltikov}}, \ and\ \bibinfo
  {author} {\bibfnamefont {F.}~\bibnamefont {Krausz}},\ }\href {\doibase
  10.1103/PhysRevLett.109.245005} {\bibfield  {journal} {\bibinfo  {journal}
  {Phys. Rev. Lett.}\ }\textbf {\bibinfo {volume} {109}},\ \bibinfo {pages}
  {245005} (\bibinfo {year} {2012})}\BibitemShut {NoStop}%
\bibitem [{\citenamefont {Dromey}\ \emph {et~al.}(2012)\citenamefont {Dromey},
  \citenamefont {Rykovanov}, \citenamefont {Yeung}, \citenamefont
  {H{\"o}rlein}, \citenamefont {Jung}, \citenamefont {Gautier}, \citenamefont
  {Dzelzainis}, \citenamefont {Kiefer}, \citenamefont {Palaniyppan},
  \citenamefont {Shah} \emph {et~al.}}]{Dromey-etal_NatPhys2012}%
  \BibitemOpen
  \bibfield  {author} {\bibinfo {author} {\bibfnamefont {B.}~\bibnamefont
  {Dromey}}, \bibinfo {author} {\bibfnamefont {S.}~\bibnamefont {Rykovanov}},
  \bibinfo {author} {\bibfnamefont {M.}~\bibnamefont {Yeung}}, \bibinfo
  {author} {\bibfnamefont {R.}~\bibnamefont {H{\"o}rlein}}, \bibinfo {author}
  {\bibfnamefont {D.}~\bibnamefont {Jung}}, \bibinfo {author} {\bibfnamefont
  {D.}~\bibnamefont {Gautier}}, \bibinfo {author} {\bibfnamefont
  {T.}~\bibnamefont {Dzelzainis}}, \bibinfo {author} {\bibfnamefont
  {D.}~\bibnamefont {Kiefer}}, \bibinfo {author} {\bibfnamefont
  {S.}~\bibnamefont {Palaniyppan}}, \bibinfo {author} {\bibfnamefont
  {R.}~\bibnamefont {Shah}},  \emph {et~al.},\ }\href {\doibase
  10.1038/nphys2439} {\bibfield  {journal} {\bibinfo  {journal} {Nature
  Physics}\ }\textbf {\bibinfo {volume} {8}},\ \bibinfo {pages} {804} (\bibinfo
  {year} {2012})}\BibitemShut {NoStop}%
\bibitem [{\citenamefont {Gonoskov}\ \emph {et~al.}(2011)\citenamefont
  {Gonoskov}, \citenamefont {Korzhimanov}, \citenamefont {Kim}, \citenamefont
  {Marklund},\ and\ \citenamefont {Sergeev}}]{Gonoskov-etal_PRE2011}%
  \BibitemOpen
  \bibfield  {author} {\bibinfo {author} {\bibfnamefont {A.~A.}\ \bibnamefont
  {Gonoskov}}, \bibinfo {author} {\bibfnamefont {A.~V.}\ \bibnamefont
  {Korzhimanov}}, \bibinfo {author} {\bibfnamefont {A.~V.}\ \bibnamefont
  {Kim}}, \bibinfo {author} {\bibfnamefont {M.}~\bibnamefont {Marklund}}, \
  and\ \bibinfo {author} {\bibfnamefont {A.~M.}\ \bibnamefont {Sergeev}},\
  }\href {\doibase 10.1103/PhysRevE.84.046403} {\bibfield  {journal} {\bibinfo
  {journal} {Phys. Rev. E}\ }\textbf {\bibinfo {volume} {84}},\ \bibinfo
  {pages} {046403} (\bibinfo {year} {2011})}\BibitemShut {NoStop}%
\bibitem [{\citenamefont {Blackburn}\ \emph {et~al.}(2018)\citenamefont
  {Blackburn}, \citenamefont {Gonoskov},\ and\ \citenamefont
  {Marklund}}]{Blackburn-etal_PRA2018}%
  \BibitemOpen
  \bibfield  {author} {\bibinfo {author} {\bibfnamefont {T.~G.}\ \bibnamefont
  {Blackburn}}, \bibinfo {author} {\bibfnamefont {A.~A.}\ \bibnamefont
  {Gonoskov}}, \ and\ \bibinfo {author} {\bibfnamefont {M.}~\bibnamefont
  {Marklund}},\ }\href {\doibase 10.1103/PhysRevA.98.023421} {\bibfield
  {journal} {\bibinfo  {journal} {Phys. Rev. A}\ }\textbf {\bibinfo {volume}
  {98}},\ \bibinfo {pages} {023421} (\bibinfo {year} {2018})}\BibitemShut
  {NoStop}%
\bibitem [{\citenamefont {Gibbon}(1996)}]{Gibbon_PRL1996}%
  \BibitemOpen
  \bibfield  {author} {\bibinfo {author} {\bibfnamefont {P.}~\bibnamefont
  {Gibbon}},\ }\href {\doibase 10.1103/PhysRevLett.76.50} {\bibfield  {journal}
  {\bibinfo  {journal} {Phys. Rev. Lett.}\ }\textbf {\bibinfo {volume} {76}},\
  \bibinfo {pages} {50} (\bibinfo {year} {1996})}\BibitemShut {NoStop}%
\bibitem [{\citenamefont {Teubner}\ and\ \citenamefont
  {Gibbon}(2009)}]{Teubner-Gibbon_RevModPhys2009}%
  \BibitemOpen
  \bibfield  {author} {\bibinfo {author} {\bibfnamefont {U.}~\bibnamefont
  {Teubner}}\ and\ \bibinfo {author} {\bibfnamefont {P.}~\bibnamefont
  {Gibbon}},\ }\href {\doibase 10.1103/RevModPhys.81.445} {\bibfield  {journal}
  {\bibinfo  {journal} {Rev. Mod. Phys.}\ }\textbf {\bibinfo {volume} {81}},\
  \bibinfo {pages} {445} (\bibinfo {year} {2009})}\BibitemShut {NoStop}%
\bibitem [{\citenamefont {Thaury}\ and\ \citenamefont
  {Qu\'{e}r\'{e}}(2010)}]{Thaury-Quere_JPhysB2010}%
  \BibitemOpen
  \bibfield  {author} {\bibinfo {author} {\bibfnamefont {C.}~\bibnamefont
  {Thaury}}\ and\ \bibinfo {author} {\bibfnamefont {F.}~\bibnamefont
  {Qu\'{e}r\'{e}}},\ }\href {\doibase 10.1088/0953-4075/43/21/213001}
  {\bibfield  {journal} {\bibinfo  {journal} {Journal of Physics B: Atomic,
  Molecular and Optical Physics}\ }\textbf {\bibinfo {volume} {43}},\ \bibinfo
  {pages} {213001} (\bibinfo {year} {2010})}\BibitemShut {NoStop}%
\bibitem [{\citenamefont {Vincenti}(2019)}]{Vincenti_PRL2019}%
  \BibitemOpen
  \bibfield  {author} {\bibinfo {author} {\bibfnamefont {H.}~\bibnamefont
  {Vincenti}},\ }\href {\doibase 10.1103/PhysRevLett.123.105001} {\bibfield
  {journal} {\bibinfo  {journal} {Phys. Rev. Lett.}\ }\textbf {\bibinfo
  {volume} {123}},\ \bibinfo {pages} {105001} (\bibinfo {year}
  {2019})}\BibitemShut {NoStop}%
\bibitem [{\citenamefont {Fedeli}\ \emph {et~al.}(2021)\citenamefont {Fedeli},
  \citenamefont {Sainte-Marie}, \citenamefont {Zaim}, \citenamefont
  {Th\'evenet}, \citenamefont {Vay}, \citenamefont {Myers}, \citenamefont
  {Qu\'er\'e},\ and\ \citenamefont {Vincenti}}]{Fedeli-etal_PRL2021}%
  \BibitemOpen
  \bibfield  {author} {\bibinfo {author} {\bibfnamefont {L.}~\bibnamefont
  {Fedeli}}, \bibinfo {author} {\bibfnamefont {A.}~\bibnamefont
  {Sainte-Marie}}, \bibinfo {author} {\bibfnamefont {N.}~\bibnamefont {Zaim}},
  \bibinfo {author} {\bibfnamefont {M.}~\bibnamefont {Th\'evenet}}, \bibinfo
  {author} {\bibfnamefont {J.~L.}\ \bibnamefont {Vay}}, \bibinfo {author}
  {\bibfnamefont {A.}~\bibnamefont {Myers}}, \bibinfo {author} {\bibfnamefont
  {F.}~\bibnamefont {Qu\'er\'e}}, \ and\ \bibinfo {author} {\bibfnamefont
  {H.}~\bibnamefont {Vincenti}},\ }\href {\doibase
  10.1103/PhysRevLett.127.114801} {\bibfield  {journal} {\bibinfo  {journal}
  {Phys. Rev. Lett.}\ }\textbf {\bibinfo {volume} {127}},\ \bibinfo {pages}
  {114801} (\bibinfo {year} {2021})}\BibitemShut {NoStop}%
\bibitem [{\citenamefont {Yi}(2021)}]{Yi_PRL2021}%
  \BibitemOpen
  \bibfield  {author} {\bibinfo {author} {\bibfnamefont {L.}~\bibnamefont
  {Yi}},\ }\href {\doibase 10.1103/PhysRevLett.126.134801} {\bibfield
  {journal} {\bibinfo  {journal} {Phys. Rev. Lett.}\ }\textbf {\bibinfo
  {volume} {126}},\ \bibinfo {pages} {134801} (\bibinfo {year}
  {2021})}\BibitemShut {NoStop}%
\bibitem [{\citenamefont {Wheeler}\ \emph {et~al.}(2012)\citenamefont
  {Wheeler}, \citenamefont {Borot}, \citenamefont {Monchoc{\'e}}, \citenamefont
  {Vincenti}, \citenamefont {Ricci}, \citenamefont {Malvache}, \citenamefont
  {Lopez-Martens},\ and\ \citenamefont
  {Qu{\'e}r{\'e}}}]{Wheeler-etal_NatPhot2012}%
  \BibitemOpen
  \bibfield  {author} {\bibinfo {author} {\bibfnamefont {J.~A.}\ \bibnamefont
  {Wheeler}}, \bibinfo {author} {\bibfnamefont {A.}~\bibnamefont {Borot}},
  \bibinfo {author} {\bibfnamefont {S.}~\bibnamefont {Monchoc{\'e}}}, \bibinfo
  {author} {\bibfnamefont {H.}~\bibnamefont {Vincenti}}, \bibinfo {author}
  {\bibfnamefont {A.}~\bibnamefont {Ricci}}, \bibinfo {author} {\bibfnamefont
  {A.}~\bibnamefont {Malvache}}, \bibinfo {author} {\bibfnamefont
  {R.}~\bibnamefont {Lopez-Martens}}, \ and\ \bibinfo {author} {\bibfnamefont
  {F.}~\bibnamefont {Qu{\'e}r{\'e}}},\ }\href {\doibase
  10.1038/nphoton.2012.284} {\bibfield  {journal} {\bibinfo  {journal} {Nature
  Photonics}\ }\textbf {\bibinfo {volume} {6}},\ \bibinfo {pages} {829}
  (\bibinfo {year} {2012})}\BibitemShut {NoStop}%
\bibitem [{\citenamefont {Vincenti}\ and\ \citenamefont
  {Qu\'er\'e}(2012)}]{Vincenti-Quere_PRL2012}%
  \BibitemOpen
  \bibfield  {author} {\bibinfo {author} {\bibfnamefont {H.}~\bibnamefont
  {Vincenti}}\ and\ \bibinfo {author} {\bibfnamefont {F.}~\bibnamefont
  {Qu\'er\'e}},\ }\href {\doibase 10.1103/PhysRevLett.108.113904} {\bibfield
  {journal} {\bibinfo  {journal} {Phys. Rev. Lett.}\ }\textbf {\bibinfo
  {volume} {108}},\ \bibinfo {pages} {113904} (\bibinfo {year}
  {2012})}\BibitemShut {NoStop}%
\bibitem [{\citenamefont {Hammond}\ \emph {et~al.}(2016)\citenamefont
  {Hammond}, \citenamefont {Brown}, \citenamefont {Kim}, \citenamefont
  {Villeneuve},\ and\ \citenamefont {Corkum}}]{Hammond-Etal_Natphot2016}%
  \BibitemOpen
  \bibfield  {author} {\bibinfo {author} {\bibfnamefont {T.~J.}\ \bibnamefont
  {Hammond}}, \bibinfo {author} {\bibfnamefont {G.~G.}\ \bibnamefont {Brown}},
  \bibinfo {author} {\bibfnamefont {K.~T.}\ \bibnamefont {Kim}}, \bibinfo
  {author} {\bibfnamefont {D.~M.}\ \bibnamefont {Villeneuve}}, \ and\ \bibinfo
  {author} {\bibfnamefont {P.~B.}\ \bibnamefont {Corkum}},\ }\href {\doibase
  10.1038/nphoton.2015.271} {\bibfield  {journal} {\bibinfo  {journal} {Nature
  Photonics}\ }\textbf {\bibinfo {volume} {10}},\ \bibinfo {pages} {171}
  (\bibinfo {year} {2016})}\BibitemShut {NoStop}%
\bibitem [{\citenamefont {Heissler}\ \emph {et~al.}(2012)\citenamefont
  {Heissler}, \citenamefont {H\"orlein}, \citenamefont {Mikhailova},
  \citenamefont {Waldecker}, \citenamefont {Tzallas}, \citenamefont {Buck},
  \citenamefont {Schmid}, \citenamefont {Sears}, \citenamefont {Krausz},
  \citenamefont {Veisz}, \citenamefont {Zepf},\ and\ \citenamefont
  {Tsakiris}}]{Heissler-etal_PRL2012}%
  \BibitemOpen
  \bibfield  {author} {\bibinfo {author} {\bibfnamefont {P.}~\bibnamefont
  {Heissler}}, \bibinfo {author} {\bibfnamefont {R.}~\bibnamefont {H\"orlein}},
  \bibinfo {author} {\bibfnamefont {J.~M.}\ \bibnamefont {Mikhailova}},
  \bibinfo {author} {\bibfnamefont {L.}~\bibnamefont {Waldecker}}, \bibinfo
  {author} {\bibfnamefont {P.}~\bibnamefont {Tzallas}}, \bibinfo {author}
  {\bibfnamefont {A.}~\bibnamefont {Buck}}, \bibinfo {author} {\bibfnamefont
  {K.}~\bibnamefont {Schmid}}, \bibinfo {author} {\bibfnamefont {C.~M.~S.}\
  \bibnamefont {Sears}}, \bibinfo {author} {\bibfnamefont {F.}~\bibnamefont
  {Krausz}}, \bibinfo {author} {\bibfnamefont {L.}~\bibnamefont {Veisz}},
  \bibinfo {author} {\bibfnamefont {M.}~\bibnamefont {Zepf}}, \ and\ \bibinfo
  {author} {\bibfnamefont {G.~D.}\ \bibnamefont {Tsakiris}},\ }\href {\doibase
  10.1103/PhysRevLett.108.235003} {\bibfield  {journal} {\bibinfo  {journal}
  {Phys. Rev. Lett.}\ }\textbf {\bibinfo {volume} {108}},\ \bibinfo {pages}
  {235003} (\bibinfo {year} {2012})}\BibitemShut {NoStop}%
\bibitem [{\citenamefont {Kando}\ \emph {et~al.}(2018)\citenamefont {Kando},
  \citenamefont {Esirkepov}, \citenamefont {Koga}, \citenamefont {Pirozhkov},\
  and\ \citenamefont {Bulanov}}]{Kando-etal_QuantBeamSci2018}%
  \BibitemOpen
  \bibfield  {author} {\bibinfo {author} {\bibfnamefont {M.}~\bibnamefont
  {Kando}}, \bibinfo {author} {\bibfnamefont {T.~Z.}\ \bibnamefont
  {Esirkepov}}, \bibinfo {author} {\bibfnamefont {J.~K.}\ \bibnamefont {Koga}},
  \bibinfo {author} {\bibfnamefont {A.~S.}\ \bibnamefont {Pirozhkov}}, \ and\
  \bibinfo {author} {\bibfnamefont {S.~V.}\ \bibnamefont {Bulanov}},\ }\href
  {\doibase 10.3390/qubs2020009} {\bibfield  {journal} {\bibinfo  {journal}
  {Quantum Beam Science}\ }\textbf {\bibinfo {volume} {2}},\ \bibinfo {pages}
  {9} (\bibinfo {year} {2018})}\BibitemShut {NoStop}%
\bibitem [{\citenamefont {Jahn}\ \emph {et~al.}(2019)\citenamefont {Jahn},
  \citenamefont {Leshchenko}, \citenamefont {Tzallas}, \citenamefont {Kessel},
  \citenamefont {Kr\"{u}ger}, \citenamefont {M\"{u}nzer}, \citenamefont
  {Trushin}, \citenamefont {Tsakiris}, \citenamefont {Kahaly}, \citenamefont
  {Kormin}, \citenamefont {Veisz}, \citenamefont {Pervak}, \citenamefont
  {Krausz}, \citenamefont {Major},\ and\ \citenamefont
  {Karsch}}]{Jahn-etal_Optica2019}%
  \BibitemOpen
  \bibfield  {author} {\bibinfo {author} {\bibfnamefont {O.}~\bibnamefont
  {Jahn}}, \bibinfo {author} {\bibfnamefont {V.~E.}\ \bibnamefont
  {Leshchenko}}, \bibinfo {author} {\bibfnamefont {P.}~\bibnamefont {Tzallas}},
  \bibinfo {author} {\bibfnamefont {A.}~\bibnamefont {Kessel}}, \bibinfo
  {author} {\bibfnamefont {M.}~\bibnamefont {Kr\"{u}ger}}, \bibinfo {author}
  {\bibfnamefont {A.}~\bibnamefont {M\"{u}nzer}}, \bibinfo {author}
  {\bibfnamefont {S.~A.}\ \bibnamefont {Trushin}}, \bibinfo {author}
  {\bibfnamefont {G.~D.}\ \bibnamefont {Tsakiris}}, \bibinfo {author}
  {\bibfnamefont {S.}~\bibnamefont {Kahaly}}, \bibinfo {author} {\bibfnamefont
  {D.}~\bibnamefont {Kormin}}, \bibinfo {author} {\bibfnamefont
  {L.}~\bibnamefont {Veisz}}, \bibinfo {author} {\bibfnamefont
  {V.}~\bibnamefont {Pervak}}, \bibinfo {author} {\bibfnamefont
  {F.}~\bibnamefont {Krausz}}, \bibinfo {author} {\bibfnamefont
  {Z.}~\bibnamefont {Major}}, \ and\ \bibinfo {author} {\bibfnamefont
  {S.}~\bibnamefont {Karsch}},\ }\href {\doibase 10.1364/OPTICA.6.000280}
  {\bibfield  {journal} {\bibinfo  {journal} {Optica}\ }\textbf {\bibinfo
  {volume} {6}},\ \bibinfo {pages} {280} (\bibinfo {year} {2019})}\BibitemShut
  {NoStop}%
\bibitem [{\citenamefont {Maier}\ \emph {et~al.}(1966)\citenamefont {Maier},
  \citenamefont {Kaiser},\ and\ \citenamefont
  {Giordmaine}}]{Maier-etal_PRL1966}%
  \BibitemOpen
  \bibfield  {author} {\bibinfo {author} {\bibfnamefont {M.}~\bibnamefont
  {Maier}}, \bibinfo {author} {\bibfnamefont {W.}~\bibnamefont {Kaiser}}, \
  and\ \bibinfo {author} {\bibfnamefont {J.~A.}\ \bibnamefont {Giordmaine}},\
  }\href {\doibase 10.1103/PhysRevLett.17.1275} {\bibfield  {journal} {\bibinfo
   {journal} {Phys. Rev. Lett.}\ }\textbf {\bibinfo {volume} {17}},\ \bibinfo
  {pages} {1275} (\bibinfo {year} {1966})}\BibitemShut {NoStop}%
\bibitem [{\citenamefont {Ping}\ \emph {et~al.}(2000)\citenamefont {Ping},
  \citenamefont {Geltner}, \citenamefont {Fisch}, \citenamefont {Shvets},\ and\
  \citenamefont {Suckewer}}]{Ping-etal_PRE2000}%
  \BibitemOpen
  \bibfield  {author} {\bibinfo {author} {\bibfnamefont {Y.}~\bibnamefont
  {Ping}}, \bibinfo {author} {\bibfnamefont {I.}~\bibnamefont {Geltner}},
  \bibinfo {author} {\bibfnamefont {N.~J.}\ \bibnamefont {Fisch}}, \bibinfo
  {author} {\bibfnamefont {G.}~\bibnamefont {Shvets}}, \ and\ \bibinfo {author}
  {\bibfnamefont {S.}~\bibnamefont {Suckewer}},\ }\href {\doibase
  10.1103/PhysRevE.62.R4532} {\bibfield  {journal} {\bibinfo  {journal} {Phys.
  Rev. E}\ }\textbf {\bibinfo {volume} {62}},\ \bibinfo {pages} {R4532}
  (\bibinfo {year} {2000})}\BibitemShut {NoStop}%
\bibitem [{\citenamefont {Cheng}\ \emph {et~al.}(2005)\citenamefont {Cheng},
  \citenamefont {Avitzour}, \citenamefont {Ping}, \citenamefont {Suckewer},
  \citenamefont {Fisch}, \citenamefont {Hur},\ and\ \citenamefont
  {Wurtele}}]{Cheng-etal_PRL2005}%
  \BibitemOpen
  \bibfield  {author} {\bibinfo {author} {\bibfnamefont {W.}~\bibnamefont
  {Cheng}}, \bibinfo {author} {\bibfnamefont {Y.}~\bibnamefont {Avitzour}},
  \bibinfo {author} {\bibfnamefont {Y.}~\bibnamefont {Ping}}, \bibinfo {author}
  {\bibfnamefont {S.}~\bibnamefont {Suckewer}}, \bibinfo {author}
  {\bibfnamefont {N.~J.}\ \bibnamefont {Fisch}}, \bibinfo {author}
  {\bibfnamefont {M.~S.}\ \bibnamefont {Hur}}, \ and\ \bibinfo {author}
  {\bibfnamefont {J.~S.}\ \bibnamefont {Wurtele}},\ }\href {\doibase
  10.1103/PhysRevLett.94.045003} {\bibfield  {journal} {\bibinfo  {journal}
  {Phys. Rev. Lett.}\ }\textbf {\bibinfo {volume} {94}},\ \bibinfo {pages}
  {045003} (\bibinfo {year} {2005})}\BibitemShut {NoStop}%
\bibitem [{\citenamefont {Ren}\ \emph {et~al.}(2007)\citenamefont {Ren},
  \citenamefont {Cheng}, \citenamefont {Li},\ and\ \citenamefont
  {Suckewer}}]{Ren-etal_NatPhys2007}%
  \BibitemOpen
  \bibfield  {author} {\bibinfo {author} {\bibfnamefont {J.}~\bibnamefont
  {Ren}}, \bibinfo {author} {\bibfnamefont {W.}~\bibnamefont {Cheng}}, \bibinfo
  {author} {\bibfnamefont {S.}~\bibnamefont {Li}}, \ and\ \bibinfo {author}
  {\bibfnamefont {S.}~\bibnamefont {Suckewer}},\ }\href {\doibase
  10.1038/nphys717} {\bibfield  {journal} {\bibinfo  {journal} {Nature
  Physics}\ }\textbf {\bibinfo {volume} {3}},\ \bibinfo {pages} {732} (\bibinfo
  {year} {2007})}\BibitemShut {NoStop}%
\bibitem [{\citenamefont {Trines}\ \emph
  {et~al.}(2011{\natexlab{a}})\citenamefont {Trines}, \citenamefont {Fi\'uza},
  \citenamefont {Bingham}, \citenamefont {Fonseca}, \citenamefont {Silva},
  \citenamefont {Cairns},\ and\ \citenamefont {Norreys}}]{Trines-etal_PRL2011}%
  \BibitemOpen
  \bibfield  {author} {\bibinfo {author} {\bibfnamefont {R.~M. G.~M.}\
  \bibnamefont {Trines}}, \bibinfo {author} {\bibfnamefont {F.}~\bibnamefont
  {Fi\'uza}}, \bibinfo {author} {\bibfnamefont {R.}~\bibnamefont {Bingham}},
  \bibinfo {author} {\bibfnamefont {R.~A.}\ \bibnamefont {Fonseca}}, \bibinfo
  {author} {\bibfnamefont {L.~O.}\ \bibnamefont {Silva}}, \bibinfo {author}
  {\bibfnamefont {R.~A.}\ \bibnamefont {Cairns}}, \ and\ \bibinfo {author}
  {\bibfnamefont {P.~A.}\ \bibnamefont {Norreys}},\ }\href {\doibase
  10.1103/PhysRevLett.107.105002} {\bibfield  {journal} {\bibinfo  {journal}
  {Phys. Rev. Lett.}\ }\textbf {\bibinfo {volume} {107}},\ \bibinfo {pages}
  {105002} (\bibinfo {year} {2011}{\natexlab{a}})}\BibitemShut {NoStop}%
\bibitem [{\citenamefont {Trines}\ \emph {et~al.}(2020)\citenamefont {Trines},
  \citenamefont {Alves}, \citenamefont {Webb}, \citenamefont {Vieira},
  \citenamefont {Fi{\'u}za}, \citenamefont {Fonseca}, \citenamefont {Silva},
  \citenamefont {Cairns},\ and\ \citenamefont
  {Bingham}}]{Trines-etal_SciRep2020}%
  \BibitemOpen
  \bibfield  {author} {\bibinfo {author} {\bibfnamefont {R.~M. G.~M.}\
  \bibnamefont {Trines}}, \bibinfo {author} {\bibfnamefont {E.~P.}\
  \bibnamefont {Alves}}, \bibinfo {author} {\bibfnamefont {E.}~\bibnamefont
  {Webb}}, \bibinfo {author} {\bibfnamefont {J.}~\bibnamefont {Vieira}},
  \bibinfo {author} {\bibfnamefont {F.}~\bibnamefont {Fi{\'u}za}}, \bibinfo
  {author} {\bibfnamefont {R.~A.}\ \bibnamefont {Fonseca}}, \bibinfo {author}
  {\bibfnamefont {L.~O.}\ \bibnamefont {Silva}}, \bibinfo {author}
  {\bibfnamefont {R.~A.}\ \bibnamefont {Cairns}}, \ and\ \bibinfo {author}
  {\bibfnamefont {R.}~\bibnamefont {Bingham}},\ }\href {\doibase
  10.1038/s41598-020-76801-z} {\bibfield  {journal} {\bibinfo  {journal}
  {Scientific reports}\ }\textbf {\bibinfo {volume} {10}},\ \bibinfo {pages}
  {1} (\bibinfo {year} {2020})}\BibitemShut {NoStop}%
\bibitem [{\citenamefont {Jang}\ \emph {et~al.}(2008)\citenamefont {Jang},
  \citenamefont {Hur}, \citenamefont {Lee}, \citenamefont {Cho}, \citenamefont
  {Namkung},\ and\ \citenamefont {Suk}}]{Jang-etal_ApplPhysLett2008}%
  \BibitemOpen
  \bibfield  {author} {\bibinfo {author} {\bibfnamefont {H.}~\bibnamefont
  {Jang}}, \bibinfo {author} {\bibfnamefont {M.~S.}\ \bibnamefont {Hur}},
  \bibinfo {author} {\bibfnamefont {J.~M.}\ \bibnamefont {Lee}}, \bibinfo
  {author} {\bibfnamefont {M.~H.}\ \bibnamefont {Cho}}, \bibinfo {author}
  {\bibfnamefont {W.}~\bibnamefont {Namkung}}, \ and\ \bibinfo {author}
  {\bibfnamefont {H.}~\bibnamefont {Suk}},\ }\href {\doibase 10.1063/1.2973395}
  {\bibfield  {journal} {\bibinfo  {journal} {Applied Physics Letters}\
  }\textbf {\bibinfo {volume} {93}},\ \bibinfo {pages} {071506} (\bibinfo
  {year} {2008})}\BibitemShut {NoStop}%
\bibitem [{\citenamefont {Cho}\ \emph {et~al.}(2014)\citenamefont {Cho},
  \citenamefont {Kim},\ and\ \citenamefont {Hur}}]{Cho-etal_ApplPhysLett2014}%
  \BibitemOpen
  \bibfield  {author} {\bibinfo {author} {\bibfnamefont {M.-H.}\ \bibnamefont
  {Cho}}, \bibinfo {author} {\bibfnamefont {Y.-K.}\ \bibnamefont {Kim}}, \ and\
  \bibinfo {author} {\bibfnamefont {M.~S.}\ \bibnamefont {Hur}},\ }\href
  {\doibase 10.1063/1.4868870} {\bibfield  {journal} {\bibinfo  {journal}
  {Applied Physics Letters}\ }\textbf {\bibinfo {volume} {104}},\ \bibinfo
  {pages} {141107} (\bibinfo {year} {2014})}\BibitemShut {NoStop}%
\bibitem [{\citenamefont {Song}\ \emph {et~al.}(2016)\citenamefont {Song},
  \citenamefont {Cho}, \citenamefont {Kim}, \citenamefont {Kang}, \citenamefont
  {Suk},\ and\ \citenamefont {Hur}}]{Song-etal_PPCF2016}%
  \BibitemOpen
  \bibfield  {author} {\bibinfo {author} {\bibfnamefont {H.~S.}\ \bibnamefont
  {Song}}, \bibinfo {author} {\bibfnamefont {M.-H.}\ \bibnamefont {Cho}},
  \bibinfo {author} {\bibfnamefont {Y.-K.}\ \bibnamefont {Kim}}, \bibinfo
  {author} {\bibfnamefont {T.}~\bibnamefont {Kang}}, \bibinfo {author}
  {\bibfnamefont {H.}~\bibnamefont {Suk}}, \ and\ \bibinfo {author}
  {\bibfnamefont {M.~S.}\ \bibnamefont {Hur}},\ }\href {\doibase
  10.1088/0741-3335/58/2/025006} {\bibfield  {journal} {\bibinfo  {journal}
  {Plasma Physics and Controlled Fusion}\ }\textbf {\bibinfo {volume} {58}},\
  \bibinfo {pages} {025006} (\bibinfo {year} {2016})}\BibitemShut {NoStop}%
\bibitem [{\citenamefont {Vieux}\ \emph {et~al.}(2013)\citenamefont {Vieux},
  \citenamefont {Ersfeld}, \citenamefont {Farmer}, \citenamefont {Hur},
  \citenamefont {Issac},\ and\ \citenamefont
  {Jaroszynski}}]{Vieux-etal_ApplPhysLett2013}%
  \BibitemOpen
  \bibfield  {author} {\bibinfo {author} {\bibfnamefont {G.}~\bibnamefont
  {Vieux}}, \bibinfo {author} {\bibfnamefont {B.}~\bibnamefont {Ersfeld}},
  \bibinfo {author} {\bibfnamefont {J.~P.}\ \bibnamefont {Farmer}}, \bibinfo
  {author} {\bibfnamefont {M.~S.}\ \bibnamefont {Hur}}, \bibinfo {author}
  {\bibfnamefont {R.~C.}\ \bibnamefont {Issac}}, \ and\ \bibinfo {author}
  {\bibfnamefont {D.~A.}\ \bibnamefont {Jaroszynski}},\ }\href {\doibase
  10.1063/1.4821581} {\bibfield  {journal} {\bibinfo  {journal} {Applied
  Physics Letters}\ }\textbf {\bibinfo {volume} {103}},\ \bibinfo {pages}
  {121106} (\bibinfo {year} {2013})}\BibitemShut {NoStop}%
\bibitem [{\citenamefont {Edwards}\ \emph {et~al.}(2017)\citenamefont
  {Edwards}, \citenamefont {Mikhailova},\ and\ \citenamefont
  {Fisch}}]{Edwards-etal_PRE2017}%
  \BibitemOpen
  \bibfield  {author} {\bibinfo {author} {\bibfnamefont {M.~R.}\ \bibnamefont
  {Edwards}}, \bibinfo {author} {\bibfnamefont {J.~M.}\ \bibnamefont
  {Mikhailova}}, \ and\ \bibinfo {author} {\bibfnamefont {N.~J.}\ \bibnamefont
  {Fisch}},\ }\href {\doibase 10.1103/PhysRevE.96.023209} {\bibfield  {journal}
  {\bibinfo  {journal} {Phys. Rev. E}\ }\textbf {\bibinfo {volume} {96}},\
  \bibinfo {pages} {023209} (\bibinfo {year} {2017})}\BibitemShut {NoStop}%
\bibitem [{\citenamefont {Derouillat}\ \emph {et~al.}(2018)\citenamefont
  {Derouillat}, \citenamefont {Beck}, \citenamefont {P\'{e}rez}, \citenamefont
  {Vinci}, \citenamefont {Chiaramello}, \citenamefont {Grassi}, \citenamefont
  {Flé}, \citenamefont {Bouchard}, \citenamefont {Plotnikov}, \citenamefont
  {Aunai}, \citenamefont {Dargent}, \citenamefont {Riconda},\ and\
  \citenamefont {Grech}}]{Smilei-paper}%
  \BibitemOpen
  \bibfield  {author} {\bibinfo {author} {\bibfnamefont {J.}~\bibnamefont
  {Derouillat}}, \bibinfo {author} {\bibfnamefont {A.}~\bibnamefont {Beck}},
  \bibinfo {author} {\bibfnamefont {F.}~\bibnamefont {P\'{e}rez}}, \bibinfo
  {author} {\bibfnamefont {T.}~\bibnamefont {Vinci}}, \bibinfo {author}
  {\bibfnamefont {M.}~\bibnamefont {Chiaramello}}, \bibinfo {author}
  {\bibfnamefont {A.}~\bibnamefont {Grassi}}, \bibinfo {author} {\bibfnamefont
  {M.}~\bibnamefont {Flé}}, \bibinfo {author} {\bibfnamefont {G.}~\bibnamefont
  {Bouchard}}, \bibinfo {author} {\bibfnamefont {I.}~\bibnamefont {Plotnikov}},
  \bibinfo {author} {\bibfnamefont {N.}~\bibnamefont {Aunai}}, \bibinfo
  {author} {\bibfnamefont {J.}~\bibnamefont {Dargent}}, \bibinfo {author}
  {\bibfnamefont {C.}~\bibnamefont {Riconda}}, \ and\ \bibinfo {author}
  {\bibfnamefont {M.}~\bibnamefont {Grech}},\ }\href {\doibase
  10.1016/j.cpc.2017.09.024} {\bibfield  {journal} {\bibinfo  {journal}
  {Comput. Phys. Commun.}\ }\textbf {\bibinfo {volume} {222}},\ \bibinfo
  {pages} {351} (\bibinfo {year} {2018})}\BibitemShut {NoStop}%
\bibitem [{\citenamefont {P\'{e}rez}\ \emph {et~al.}(2012)\citenamefont
  {P\'{e}rez}, \citenamefont {Gremillet}, \citenamefont {Decoster},
  \citenamefont {Drouin},\ and\ \citenamefont {Lefebvre}}]{Perez-etal_PoP2012}%
  \BibitemOpen
  \bibfield  {author} {\bibinfo {author} {\bibfnamefont {F.}~\bibnamefont
  {P\'{e}rez}}, \bibinfo {author} {\bibfnamefont {L.}~\bibnamefont
  {Gremillet}}, \bibinfo {author} {\bibfnamefont {A.}~\bibnamefont {Decoster}},
  \bibinfo {author} {\bibfnamefont {M.}~\bibnamefont {Drouin}}, \ and\ \bibinfo
  {author} {\bibfnamefont {E.}~\bibnamefont {Lefebvre}},\ }\href {\doibase
  10.1063/1.4742167} {\bibfield  {journal} {\bibinfo  {journal} {Phys. of
  Plasmas}\ }\textbf {\bibinfo {volume} {19}},\ \bibinfo {pages} {083104}
  (\bibinfo {year} {2012})}\BibitemShut {NoStop}%
\bibitem [{Note1()}]{Note1}%
  \BibitemOpen
  \bibinfo {note} {The spatial resolution was doubled, $\Delta {x}\simeq
  0.34{\protect \rm \protect \tmspace +\thinmuskip {.1667em}nm}$, and the
  number of particles per cell was increased to 8000 per cell and per species,
  i.e., in total 16 times the number of particles compared with the nominal
  simulations. We deemed that the simulation results of the spontaneous RS were
  sufficiently converged, when the energy difference between 4000 and 8000
  particles per cell was smaller than 10\protect \tmspace +\thinmuskip
  {.1667em}\%.}\BibitemShut {Stop}%
\bibitem [{\citenamefont {Max}\ \emph {et~al.}(1974)\citenamefont {Max},
  \citenamefont {Arons},\ and\ \citenamefont {Langdon}}]{Max-etal_PRL1974}%
  \BibitemOpen
  \bibfield  {author} {\bibinfo {author} {\bibfnamefont {C.~E.}\ \bibnamefont
  {Max}}, \bibinfo {author} {\bibfnamefont {J.}~\bibnamefont {Arons}}, \ and\
  \bibinfo {author} {\bibfnamefont {A.~B.}\ \bibnamefont {Langdon}},\ }\href
  {\doibase 10.1103/PhysRevLett.33.209} {\bibfield  {journal} {\bibinfo
  {journal} {Phys. Rev. Lett.}\ }\textbf {\bibinfo {volume} {33}},\ \bibinfo
  {pages} {209} (\bibinfo {year} {1974})}\BibitemShut {NoStop}%
\bibitem [{\citenamefont {Weber}\ \emph {et~al.}(2013)\citenamefont {Weber},
  \citenamefont {Riconda}, \citenamefont {Lancia}, \citenamefont {Marqu\`es},
  \citenamefont {Mourou},\ and\ \citenamefont {Fuchs}}]{Weber-etal_PRL2013}%
  \BibitemOpen
  \bibfield  {author} {\bibinfo {author} {\bibfnamefont {S.}~\bibnamefont
  {Weber}}, \bibinfo {author} {\bibfnamefont {C.}~\bibnamefont {Riconda}},
  \bibinfo {author} {\bibfnamefont {L.}~\bibnamefont {Lancia}}, \bibinfo
  {author} {\bibfnamefont {J.-R.}\ \bibnamefont {Marqu\`es}}, \bibinfo {author}
  {\bibfnamefont {G.~A.}\ \bibnamefont {Mourou}}, \ and\ \bibinfo {author}
  {\bibfnamefont {J.}~\bibnamefont {Fuchs}},\ }\href {\doibase
  10.1103/PhysRevLett.111.055004} {\bibfield  {journal} {\bibinfo  {journal}
  {Phys. Rev. Lett.}\ }\textbf {\bibinfo {volume} {111}},\ \bibinfo {pages}
  {055004} (\bibinfo {year} {2013})}\BibitemShut {NoStop}%
\bibitem [{\citenamefont {Trines}\ \emph
  {et~al.}(2011{\natexlab{b}})\citenamefont {Trines}, \citenamefont {Fiuza},
  \citenamefont {Bingham}, \citenamefont {Fonseca}, \citenamefont {Silva},
  \citenamefont {Cairns},\ and\ \citenamefont
  {Norreys}}]{Trines-etal_NatPhys2011}%
  \BibitemOpen
  \bibfield  {author} {\bibinfo {author} {\bibfnamefont {R.~M. G.~M.}\
  \bibnamefont {Trines}}, \bibinfo {author} {\bibfnamefont {F.}~\bibnamefont
  {Fiuza}}, \bibinfo {author} {\bibfnamefont {R.}~\bibnamefont {Bingham}},
  \bibinfo {author} {\bibfnamefont {R.~A.}\ \bibnamefont {Fonseca}}, \bibinfo
  {author} {\bibfnamefont {L.~O.}\ \bibnamefont {Silva}}, \bibinfo {author}
  {\bibfnamefont {R.~A.}\ \bibnamefont {Cairns}}, \ and\ \bibinfo {author}
  {\bibfnamefont {P.~A.}\ \bibnamefont {Norreys}},\ }\href {\doibase
  10.1038/nphys1793} {\bibfield  {journal} {\bibinfo  {journal} {Nature
  Physics}\ }\textbf {\bibinfo {volume} {7}},\ \bibinfo {pages} {87} (\bibinfo
  {year} {2011}{\natexlab{b}})}\BibitemShut {NoStop}%
\bibitem [{\citenamefont {Sundstr\"{o}m}\ \emph
  {et~al.}(2020{\natexlab{a}})\citenamefont {Sundstr\"{o}m}, \citenamefont
  {Gremillet}, \citenamefont {Siminos},\ and\ \citenamefont
  {Pusztai}}]{ElectronPaper2020}%
  \BibitemOpen
  \bibfield  {author} {\bibinfo {author} {\bibfnamefont {A.}~\bibnamefont
  {Sundstr\"{o}m}}, \bibinfo {author} {\bibfnamefont {L.}~\bibnamefont
  {Gremillet}}, \bibinfo {author} {\bibfnamefont {E.}~\bibnamefont {Siminos}},
  \ and\ \bibinfo {author} {\bibfnamefont {I.}~\bibnamefont {Pusztai}},\ }\href
  {\doibase 10.1017/S0022377820000264} {\bibfield  {journal} {\bibinfo
  {journal} {Journal of Plasma Physics}\ }\textbf {\bibinfo {volume} {86}},\
  \bibinfo {pages} {755860201} (\bibinfo {year}
  {2020}{\natexlab{a}})}\BibitemShut {NoStop}%
\bibitem [{\citenamefont {Sundstr\"{o}m}\ \emph
  {et~al.}(2020{\natexlab{b}})\citenamefont {Sundstr\"{o}m}, \citenamefont
  {Gremillet}, \citenamefont {Siminos},\ and\ \citenamefont
  {Pusztai}}]{IonPaper2020}%
  \BibitemOpen
  \bibfield  {author} {\bibinfo {author} {\bibfnamefont {A.}~\bibnamefont
  {Sundstr\"{o}m}}, \bibinfo {author} {\bibfnamefont {L.}~\bibnamefont
  {Gremillet}}, \bibinfo {author} {\bibfnamefont {E.}~\bibnamefont {Siminos}},
  \ and\ \bibinfo {author} {\bibfnamefont {I.}~\bibnamefont {Pusztai}},\ }\href
  {\doibase 10.1088/1361-6587/ab9a62} {\bibfield  {journal} {\bibinfo
  {journal} {Plasma Physics and Controlled Fusion}\ }\textbf {\bibinfo {volume}
  {62}},\ \bibinfo {pages} {085015} (\bibinfo {year}
  {2020}{\natexlab{b}})}\BibitemShut {NoStop}%
\end{thebibliography}%
\end{document}